\documentclass{revtex4}
\usepackage{graphicx}
\usepackage{amsmath}
\usepackage{color}

\begin{document}

\title{Could trapped quintessence account for the laser-detuning-dependent acceleration of cold atoms in varying-frequency time-of-flight experiments?}

\author{Hai-Chao Zhang}
\email{zhanghc@siom.ac.cn}

\author{Xin-Ping Xu}
\author{Jing-Fang Zhang}
\author{Chuan Wang}

\affiliation{Key Laboratory for Quantum Optics, Shanghai Institute of Optics and Fine Mechanics, Chinese Academy of Sciences, Shanghai 201800, China }

\date{\today}

\begin{abstract}
Using a trapped quintessence model, a series of time-of-flight (TOF) experiments with a different frequency of probe light were designed and performed. The varying-frequency TOF (VFTOF) experiments demonstrated that the fall acceleration of test atoms is dependent on the detuning of the probe light frequency with respect to the atomic transition frequency. In appropriately designed experiments, if the scalar field in the model accounts for the accelerated expansion of the Universe entirely, the field will result in an observable fifth force. Meanwhile, the trapped quintessence model still satisfies all experimental bounds on deviations from general relativity due to both the saturation effect and the short interaction range of the scalar field. The scalar saturates at a value corresponding to the cosmological constant when the microscopic nonrelativistic matter density is large enough. The interaction range of the scalar is inversely proportional to the square root of the microscopic nonrelativistic matter density. The interaction range has been estimated to be several $\mu \rm{m}$ in the current cosmic density $\sim 10^{-27}\,\rm{kg/m^3}$. The Universe is assumed to be permeated with fuzzy dark matter, which means that the microscopic nonrelativistic matter density defined through the quantum wavefunctions of the ultralight particles can be used on the cosmic scale.

In an almost completely empty space between atoms of a dilute atomic gas in an ultrahigh vacuum chamber, the interaction range of the scalar field may approach to the order of $\sim 1\,\mu \rm{m}$ in the presence of dark matter and then the scalar field might be detected in laboratories. Since the trapped quintessence model hypothesizes that the scalar strongly couples to nonrelativistic matter but cannot couple to radiation, the source for generating the fifth force was experimentally set up by the laser-irradiated background atoms in the ultrahigh vacuum chamber. The mass density of the source was altered by detuning the frequency of the probe laser light from the atomic resonance transition. The test atoms were prepared by the laser cooling technique and located initially above the probe light. When the test atoms were released from their initial positions, they were able to pass through the region of the source that generated the fifth force to be measured. Thus, if the scalar field existed, the corresponding fifth force might be sensed by the test atoms even if the interaction range was extremely short. By measuring the fall acceleration of the test atoms with the TOF method step-by-step in the detuning frequency domain of the probe light, we derived the dispersion curves of the measured acceleration versus the frequency detuning of the probe light. When the nonrelativistic matter density of the source increased due to the energy gained from the laser light, the test atoms were pulled to the center of the source, and vice versa. If the trapped quintessence model is correct, the observed detuning-dependent acceleration in the VFTOF scheme suggests a closed Universe, i.e., a positive spatial curvature Universe.
\end{abstract}

\pacs{
PACS 04.80.Cc -Experimental tests of gravitational theories
PACS 95.36.+x -Dark energy
PACS 04.50.Kd -Modified theories of gravity
PACS 42.50.Ct -Quantum description of interaction of light and matter; related experiments
PACS 42.50.Vk -Mechanical effects of light on atoms, molecules, electrons, and ions   }
\keywords{dark energy, fifth force, cosmological constant, symmetry-breaking, quintessence, scalar field, time-of-flight}
\maketitle

\section{Introduction}\label{Introduction}

It is well known that there exist four fundamental interactions in nature: the gravitational, electromagnetic, strong and weak interactions. The first two produce long-range forces and the other produce forces at subatomic distances. These fundamental interactions are used to explain almost all physical and chemical phenomena. However, an accelerated expanding Universe observed in astronomy \cite{zn1,z27,z28,z1,z1i,z78,z64,z87,z65} implies that there might exist another so-called `fifth interaction' in nature \cite{z2,z3,z4,z5,z6,z7,z45,z58}. The cosmic acceleration is currently accounted by dark energy having the distinctive feature of a negative pressure but there is no agreement on the origin of dark energy \cite{zhcn59}. One of the most competitive schemes for the origin introduces a new scalar field into Einstein's general relativity \cite{z20,z29} to characterize dark energy \cite{z34,z9,z11,z112}. According to quantum field theory, the scalar field should produce fifth forces as long as the scalar couples with ordinary matter \cite{z33,z36,z37,z38,z39,z40}. If the coupled scalar can account for the cosmic acceleration entirely, then the following question arises: 1. Why have not fifth forces been detected in laboratory and the solar-system to date \cite{z8,z21,z46,z43}? 2. How would fifth forces be detected if they existed \cite{z43,z46,z44,z50,z49}?

The coupled scalar field theories of dark energy, such as chameleon-like models \cite{z3,z4,z5,z6,z7}, explain the absence of the observable interaction via screening mechanisms. The mechanisms state that the forces might be strong in a thin-shell region near the surface of a source object but greatly suppressed inside and outside of the source object because both the interaction ranges and the strengths of the forces depend on the ambient matter density. The bigger the ambient matter density, the weaker the fifth forces. Since most laboratory experiments are carried out in high matter density cases, the forces arising from the scalar are strongly suppressed. Unfortunately, chameleon no-go theorems indicate that chameleon-like scalar field cannot account for the cosmic acceleration except as some form of dark energy \cite{z13,z101,z57}. In other words, it is regarded that the fifth forces are unable to be suppressed to satisfy local tests of gravity if dark energy is able to be driven entirely by the scalar field. Consequently, the expression of the fifth forces cannot be specifically determined due to the lack of the strict cosmological constraints. Correspondingly, the design parameters for laboratory-based test of the forces are difficult to be estimated. A few precision experiments for probing the scalar fifth forces have been performed in ultrahigh-vacuum situation as required by chameleon-like models \cite{z102,z47,z80}. No evidence in these experiments indicates the presence of the fifth force but these ``zero" results can be used to constrain the parameters of the chameleon-like models \cite{z15}.

By introducing a broken-symmetry interaction between matter and the scalar, we have demonstrated that the so-called trapped quintessence can account for the cosmic accelerated expansion entirely \cite{zhc1}. The scalar field is trapped by one of the broken-symmetry interaction potential wells and then its self-interaction potential value at the well bottom mimics the cosmological constant. As suggested by the literature \cite{randr5,randr15,randr16}, the three ${\mathbf{Z}_{\rm{2}}}$ ($\phi  \to  - \phi $) symmetry-related models \cite{z6,randr3,randr4} are now well-known as the symmetron models. Based on the ${\mathbf{Z}_{\rm{2}}}$ symmetry, Pietroni has defined a scalar-tensor theory and used his theory to explore the cosmological consequences \cite{randr3}. Applying the couplings of the scalar field to matter much stronger than gravitational field, the dependence of masses and the fine-structure constant on the ambient matter density was explored by Olive and Pospelov \cite{randr4}. To satisfy all local tests of gravity, a similar framework such as \cite{z6} was proposed by Hinterbichler and Khoury. It allows the scalar to mediate a long-range ($\sim \rm{Mpc}$) fifth force \cite{z6}. As a matter of fact, before 1998 when the accelerating expansion of the Universe was not discovered, symmetron-related theories \cite {randr12,randr13,randr14} were reported to modify general relativity (GR). Since our theoretical model relies on the ${\mathbf{Z}_{\rm{2}}}$ symmetry and its spontaneous breaking, the model is considered to be a concrete example of the symmetron mechanism. To make the quintessence in our model account for the cosmic accelerated expansion entirely, the quintessence potential combined with the interaction potential between matter and the scalar are carefully chosen to satisfy the adiabatic tracking condition \cite{zhc1}. Because a cosmic model delineates the expanding Universe, the requirement that the scalar should be trapped tightly and sit stably at one of the bottoms of the effective potential is a key distinction from \cite{z6,randr3,randr4,randr12,randr13,randr14}.

The trapped quintessence model also differs from the original simplest quintessence model \cite{z57} in that there is no nongravitational interactions of the scalar field with other sectors \cite{zhcn59}. Simultaneously, the fifth force produced by the coupled scalar has an extremely short interaction range ($\sim \mu \rm{m}$) and a very small peak strength. Therefore, it is almost unobservable in the common laboratory experiments and the solar-system tests of local gravity. In addition, the value of the coupled scalar at the potential well bottom approaches a fixed number when the matter density is large enough. This means that the magnitude of the fifth force generated by a source saturates at a certain value even if the mass density of the source continuously increase \cite{zhc1}. We call the phenomenon saturation effect of the coupled scalar field. The ``zero" results of precision experiments \cite{z102,z47,z80} can be explained by the saturation effect, the short interaction range of the scalar \cite{z66}, and the relatively distant sensors away from the sources. In order to detect fifth forces described by the trapped quintessence scheme, the probe experiment might be best to allow that test particle is able to travel through the surface of source since the fifth force appears in the very narrow neighborhood of the surface.

In our experimental research, we are only going to evaluate the possibility of the existence of the fifth force rather than to measure accurately the parameters of the fifth force. An ultrahigh-vacuum chamber containing very dilute atomic gas was used to simulate the low-density conditions of the cosmic empty space. Both the source object for producing the fifth force and the test object for probing the fifth force were formed from the atoms of the dilute atomic gas. Since both are dilute atomic clouds, they could pass through each other. Thus, the fifth force that might be concealed in the thin-shell region near the source's surface could be sensed. The test object was produced by the optical molasses technique \cite{zexn10}, which indeed is an atomic ensemble sample containing millions of non-interacting identical atoms. Based on the principle of ensemble detection, a single experiment with so many of these non-interacting atoms is equivalent to the accumulated results of successive single-atom testing experiment repeated many times.

Since the interaction range of the scalar is strongly dependent on the \emph{microscopic} matter density, it would approach infinity in the large void regions between atoms if there is no fuzzy dark matter filled, and it would be extremely short at atom locations where the matter density peaks the most. However, this ``billiard balls" description is only valid for atoms with high velocities in a weakly interacting gas. For atoms with velocities approaching zero, the atoms must be regarded as quantum-mechanical wavepackets which extend on the order of a de Broglie wavelength $\lambda_{\text{de}}=h/(Mv)$ with $h$, $M$ and $v$ being the Planck constant, the particle mass and velocity, respectively. $\lambda_{\text{de}}$ can be used to represent the position uncertainty associated with the velocity of atom. In general, the average distance between atoms $ n^{-1/3} $ with $n$ being the atomic number density is much larger than the size of the atoms which is characterized by the s-wave scattering length \cite{zhce16}; but it is smaller than the mean free path in the gas. The collision between atoms is negligible. Since the exact microscopic matter density is defined by particle wavefunction, we will introduce an effective background density (EBD) rather than an average density of the source atomic gas. When the de Broglie wavelength is comparable to or larger than the interatomic separation, the atomic wavepackets ``overlap" and an almost homogeneous density distribution forms in the gas. This homogeneous matter density of EBD can be regarded as the initial microscopic matter density of the source object. When the matter density changes locally, the inhomogeneous source object is generated in the gas and then induces the variation of the scalar field. Detailed description of the source object is in the main text.

The region of the source gas in our experiment was covered by a pair of counter-propagating laser beams. Because the scalar field is assumed to couple to nonrelativistic matter rather than to radiation, the nonrelativistic mass density of the source can be adjusted by tuning the frequency of the laser beams based on the energy exchange between laser photons and narelativistic atoms. Thus, our experimental setting may also be used to demonstrate indirectly that the scalar does not couple to photons.

The article is organized as follows. In Sec. \ref{model}, we briefly describe the trapped quintessence model whose parameters have been determined by the astrophysical observations. The trapped quintessence field should generate the fifth force due to the inhomogeneous distribution of matter. To accommodate our experiment scheme for testing the fifth force, a cylindrical source embedded in the medium of background uniform density is analyzed specifically in this section. In Sec. \ref{DESIGN}, we discuss our probing strategy for the scalar fifth force and the corresponding experimental design. We show that the advantage of the VFTOF method using cold atoms to detect the fifth force that is generated by the laser-controlled source. In Sec. \ref{RESULT}, we show the experimental results of the measured acceleration of cold atoms versus the laser detuning and explain the measured dispersion curves by using the trapped quintessence model. Finally, in Sec. \ref{Conclusions}, we present further discussion and the conclusion.

\section{THE THEORETICAL MODEL }\label{model}

By invoking a broken-symmetry interaction potential between the scalar field and matter, the scalar with its self-interaction potential can entirely account for the cosmic accelerated expansion \cite{zhc1}. Differing from the original simplest quintessence model \cite{z57}, the coupling of the scalar to matter must result in the scalar fifth force.

\subsection{The trapped quintessence model }\label{trappedquin}
For a static, space-variable density source, the dynamics of the scalar field $\phi $ with the dimension of energy are governed by an effective potential density ${V_{{\rm{eff}}}}(\phi ){\rm{ }}$ \cite{z2,zhc1}, i. e.,
\begin{equation}
{\hbar ^2}{c^2}{\nabla ^2}\phi  = {V_{{\rm{eff,}}\phi }}(\phi ),\label{ex1}
\end{equation}
where $\hbar $ is the reduced Planck constant, $c$ is the speed of light and the effective potential density ${V_{{\rm{eff}}}}\left( \phi  \right){\rm{  =  }}V(\phi ){\rm{  +  }}{V_{{\mathop{\rm int}} }}$ which is a sum of a self-interaction $V(\phi )$ and an interaction ${V_{{\rm{int}}}}$ with matter. We will use commas to denote derivatives, e.g., ${V_{,\phi }}{\rm{  = }}dV(\phi )/d\phi $. The interaction potential density is an explicit function of local matter density $\varrho $ \cite{z2,zhc1},
\begin{equation}
{V_{{\mathop{\rm int}} }}{\rm{  =  }}{\varrho} {\hbar ^3}{c^5}\left[ {{A^{{\rm{1 - 3}}w}}\left( \phi  \right) - 1} \right], \label{ex2}
\end{equation}
where $w$ is the equation of state given by $w = P/(\varrho {c^2})$ with pressure $P$, $A\left( \phi  \right)$ is the coupling function between the scalar and matter. For non-relativistic matter the equation of state $w = 0$, and for radiation $w = 1/3$ meaning that the scalar field does not couple to radiation. In order to mimic the cosmological constant, it has been demonstrated that the coupling function $A\left( \phi  \right)$ and the self-interaction potential $V(\phi )$ can be chosen as follows \cite{zhc1}
\begin{subequations}\label{equ16revise}
\begin{eqnarray}
\nonumber A\left( \phi  \right) &=& \exp \left[ {\frac{{{{\left( {{\phi ^2} - M_2^2{c^4}} \right)}^2}}}{{4M_1^4{c^8}}}} \right]\\
&\approx & 1 + \frac{1}{{4M_1^4{c^8}}}{\left( {{\phi ^2} - M_2^2{c^4}} \right)^2},\label{ex3a}\\
V(\phi ) &=& \frac{\lambda }{{\rm{4}}}{\phi ^4},\label{ex3b}
\end{eqnarray}
\end{subequations}
where ${M_1}$, ${M_2}$ are parameters with mass dimension and $\lambda $ is a dimensionless parameter.

Differing from the traditional slow-roll quintessence models \cite{z57,zhcn59}, the coupled scalar here is clamped at one of the minima of the effective potential density. It has been pointed out in \cite{zhc1} that the trapped quintessence model must include a component of fuzzy dark matter with ultralight mass \cite{z67,z108} so as to obtain the correct cosmological constant. The spatial extension property of the wavepackets of ultralight particles guarantees that the cosmic space is filled with dark matter everywhere. Therefore, the possibility of appearing observable long-range scalar fifth force \cite{z66} is avoided since the matter-density-dependent interaction range of the scalar is extremely short \cite{zhc1,zhc2}. For the current cosmic density $\sim 10^{-27}\,\rm{kg/m^3}$, the interaction range is estimated to be several $\mu \rm{m}$. The matter density $\varrho$ in Eq. (\ref{ex2}) should be defined through the wavefunctions of particles including dark matter and atomic matter. Thus, $\varrho$ is essentially a quantum physical quantity and should be called the \emph{microscopic} matter density to emphasize the quantum wavefunction properties of fuzzy dark matter particles although this concept is now used on the \emph{cosmic scale}.

In principle, the three parameters $\lambda$, $M_1$ and $M_2$ can be determined by the current astronomical observation data, e.g., the cosmological constant, the Hubble constant, the ratio of the matter density to the total density of the Universe, and the transition time of the Universe expansion from deceleration to acceleration. In our implementation, assuming that the values of ${M_1}$ and ${M_2}$ on the order of the mass scale of the cosmical constant, we can infer that the value of $\lambda$-parameter falls on the order $\mathcal{O}(1)$ based on the equations (22) and (27) of the literature \cite{zhc1}. Then, under the cosmic constraints, we impose that the chosen values of the three parameters can correspond to an observable fifth force in laboratory, so that the model can be tested experimentally in the fixed parameters. We find that, when the self-interaction potential is $V(\phi )=\phi ^4/4!$ which corresponds to $\lambda  = 1/6$ \cite{z37,zhc1}, a relatively larger Compton wavelength of $\mu \rm{m}$ of the scalar field can be obtained in the current cosmic density based on the equation (22) of the literature \cite{zhc1}. Since the $\mu \rm{m}$-scale interaction range is large enough, it is not very difficult to design an experiment to probe the scalar fifth force in laboratory. Meanwhile, compared to the size of the Solar system, the $\mu \rm{m}$-scale interaction range is considerably shorter and the theoretical model satisfies the Solar system tests of gravity.

When the acceptable value $\lambda = 1/6 $ is fixed, the values of ${M_1}$ and ${M_2}$ are sensitive to the spacial curvature of the Universe. For an almost flat universe shown in literature \cite{z1,z1i}, the ratio ${M_2}/{M_1}$ has been demonstrated in literature \cite{zhc1} to be larger than \emph{four}. Since ${M_2}c^2$ is the minimum of $A\left( \phi  \right)$, the value of the self-interaction potential at the minimum is $({M_2}c^2 )^4/4!$. Thus, when the matter density $\varrho$ in Eq. (\ref{ex2}) is large enough, ${M_2}c^2(4!)^{-1/4}$ can be roughly used to denote the cosmological constant \cite{zhc1}. Consequently, we often use the value of ${M_2}$ and the ratio of ${M_1}/{M_2}$ to discuss the topic in the following text. As a typical example for the almost flat case \cite{zhc1}, the parameters are shown in Eq. (\ref{ex4}):
\begin{equation}
 {M_2} = 4.96168 \; {\rm{meV/}}{c^2}=8.845\times {{10}^{ - 39}}\,{\rm{kg}},\, {M_{1}} = \frac{M_{2}}{8}.\label{ex4}
\end{equation}
If the Universe is closed as show in \cite{z114}, the parameters are determined as follows \cite{zhc2}
\begin{equation}
M_{2} = 4.40353{\text{ meV/}}{c^2}=7.850\times {{10}^{ - 39}}\,{\rm{kg}},\, {M_1} = \frac{{{M_2}}}{3}.\label{ex5}
\end{equation}

It is difficult to determine the truth-value of $\lambda$ as well as the related truth-values of $M_1$ and $M_2$, and if the broken-symmetry model reflects the nature of the Universe. The concrete values of the presented results in this paper are based on the special selection of $\lambda=1/6$. However, the general discussion is independent of the special selection. Since $\lambda=1/6$ falls on the possible values of the $\mathcal{O}(1)$ order for $\lambda$, one might expect that the values of the presented results would be very close to their truth-values if the corresponding quantities indeed existed. It can be seen in the following that the fifth force would be larger in a closed universe than that in a flat universe due to the sensitivity of the interaction range of the scalar to the ratio of ${M_1}/{M_2}$. Therefore, with the measurement precision improving, the nature of the spacial curvature of the Universe could emerge prominently in laboratory-based tests of the scalar fifth force.

Our trapped quintessence model needs fuzzy dark matter. Dark matter \cite{randr8,randr17} is presumed from the global gravitational effect including the cosmic microwave background (CMB) and flat galactic rotation curves. Although the dark matter hypothesis is now generally accepted, it faces a number of challenges. For example, dismissing dark matter, the flat rotation curves of galaxies can be explained by not only the modified Newtonian dynamics (MOND) \cite{randr7} but also the symmetron scalar field theories\cite{randr13,randr15}. Consequently, if dark matter existed, the mass density of dark matter near an astronomical object, such as our Earth, would not be definitively determined. Possibilities of the mass density of dark matter near the Earth's surface range from zero \cite{randr7,randr13,randr15} to several hundred-thousand times of the current cosmic density of $\sim 10^{-27}\,\rm{kg/m^3}$ in literature \cite{randr17,randr9,ranr18}. No dark matter particles have been detected directly in laboratories. The particle nature of dark matter as well as the kinds of the particles in it have not been determined. The possibilities of the mass of dark matter candidate range considerably large in literature \cite{randr16}, that is, from $\sim 10^{22} \, \rm{eV}$ \cite{randr10} to $\sim 10^{2} \, M_{\odot}$ \cite{randr11}.

\subsection{The fifth force in the trapped quintessence model }\label{forceintrappedquin}

For the source of a static non-relativistic matter distribution, apart from the Newtonian force, the fifth force on a test particle due to the scalar field is \cite{z2,z39,z40,z13}:
\begin{equation}
\vec {a} =  - {c^2} \nabla \ln A\left( \phi  \right).\label{ex6}
\end{equation}

In response to the experimental situation of our VFTOF, it is necessary to distinguish the homogeneous density background from the density-space-variable source that generates the spatial variation of the scalar field. Let us imagine the source embedded in the medium of background uniform microscopic density ${\rho _{\rm{b}}}$, i.e. EBD $\equiv{\rho _{\rm{b}}}$. Apparently, the EBD, just a smooth part of the total matter density $\varrho$ in Eq. (\ref{ex2}), should include not only the contribution of atomic matter but also the contribution of dark matter. However, to simplify the analysis of the experimental scheme, we focus only on the contribution of atomic matter to the EBD. The contribution of dark matter to the EBD can be regarded as a thinner background density in the trapped quintessence model with fuzzy dark matter. Unless otherwise stated, the contribution of dark matter is ignored in the following analysis of the experimental probe mechanism. The detailed description is presented in Appendix \ref{appA}.

In an ultrahigh-vacuum chamber for performing the VFTOF experiment, we classify the working atoms into two types: (1) ``billiard balls" type which can be described classically both for collisions and free movement of atoms, (2) extension wavepackets type which must be described by quantum mechanics.

For the ``billiard balls" type, the microscopic matter density is strongly peaked at the locations of the individual atoms, and the separation distances between the atoms are much larger than their radii (characterized by the s-wave scattering length \cite{zhce16}). Therefore, the microscopic matter density is zero in the spaces between the ``billiard-ball-like" atoms. Of course, the motions of the billiard-ball-like atoms can also be described by quantum mechanics. If the wavepacketes of atoms are so narrow that the local density of a single atom can be regarded as a $\delta$ function of the space point, the classical description is a good approximation. Unlike the microscopic matter density, an average mass density ${\rho_{\rm{c}}}=n M$ with atomic number density $n$ and atomic mass $M$, is often used to describe vacuum degree in the scale of the vacuum chamber size. The average mass density ${\rho_{\rm{c}}}$ can be called as \emph{macroscopic} matter density and is not interesting in this paper since the motion of the scalar is related to the \emph{microscopic} matter density rather than the macroscopic matter density.

For the extension wavepackets type, we use the EBD ${\rho _{\rm{b}}}$ to describe the microscopic matter density of the background. The EBD ${\rho _{\rm{b}}}$ is defined for the background by the overlapping extension wavefunctions of the atoms. The largest extension scale of the wavepackets is characterized by the size of the vacuum chamber. It should be emphasized that this microscopic background density is very different form the average density ${\rho_{\rm{c}}}$ which reflects the vacuum degree. In other words, the locally peaking part of the microscopic matter density that reflects the locality of the sparsely dotted and floated individual billiard-ball-like atoms is excluded in the definition of the background density since the spatial volume occupied by the billiard-ball-like atoms can be ignored in the dilute gas. The necessity of introducing EBD will be further discussed in Sec. \ref{DESIGN}. In this homogeneous background, the scalar field satisfies that ${\nabla ^2}\phi {\rm{  = 0}}$ and the equilibrium value of the scalar field is ${\phi _{\rm{b}}}{\rm{ = }}{\phi _{\min }}\left( {{\rho _{\rm{b}}}} \right)$ corresponding to a minimum of the effective potential density of ${V_{{\rm{eff}}}}(\phi ){\rm{ }}$.

Assuming that the source's matter density $\rho  = {\rho _{\rm{b}}} + \rho_{\rm{\delta} }$ and $|\rho_{\delta}| \ll \rho _{\rm{b}}$, we can expand the scalar field $\phi$ around the background value ${\phi _{\rm{b}}}$, i.e., $\phi  = {\phi _{\rm{b}}} + \phi_{\delta} $. Notice that, unlike Eq. (\ref{ex2}) in which the matter density is denoted by $\varrho$, the source's matter density here is denoted by $\rho$ so as to indicate that the billiard-ball-like atoms are \emph{not} included in it. How to obtain $\rho_{\rm{\delta}} $ experimentally will be discussed in Sec. \ref{DESIGN}. Thus, an equation of motion for a massive scalar field $\phi_{\delta} $ is obtained from Eq. (\ref{ex1}) as follows:
\begin{equation}
\left( {{\nabla ^2} - \frac{{m_{{\rm{eff}}}^{\rm{2}}{c^2}}}{{{\hbar ^2}}}} \right)\phi_{\delta} = {A_{,\phi }}\left( {{\phi _{\rm{b}}}} \right)\hbar {c^{\rm{3}}} \rho_{\delta} ,\label{ex7}
\end{equation}
where the effective mass $m_{{\rm{eff}}}^{\rm{2}} \equiv {V_{{\rm{eff}}}}_{,\phi \phi }\left( {{\phi _{\rm{b}}}} \right)/{c^4} = {{{\rm{2}}\rho _{\rm{b}} {\hbar ^3}{M_2}^{\rm{2}}}}/({{{M_1}^{\rm{4}}{c^3}}})$ depends on the EBD ${\rho _{\rm{b}}}$ \cite{zhc1}. Thus, the interaction range of the scalar described by the scalar's Compton wavelength ${\mathchar'26\mkern-10mu\lambda _{\rm{c}}} \equiv \hbar /\left( {{m_{{\rm{eff}}}}c} \right)$ is naturally EBD-dependent. Obviously, the interaction range calculated by using the parameters shown in Eq. (\ref{ex4}) is slightly smaller than that calculated by using the parameters shown in Eq. (\ref{ex5}). The scalar fifth force exerted on a test particle shown in Eq. (\ref{ex6}) can be rewritten as
\begin{equation}
\vec {a} =  - {c^2}\frac{{{A_{,\phi }}\left( {{\phi _{\rm{b}}}} \right)}}{{A\left( {{\phi _{\rm{b}}}} \right)}}\nabla \phi_{\delta} .\label{ex8}
\end{equation}

Assuming that the source is cylindrical and extends infinitely in the axis direction while the radial position is $\vec r \equiv (x, y)$ with the origin of coordinates at the axis of the cylinder, the solution to Eq. (\ref{ex7}) can then be expressed as
\begin{equation}
 \phi_{ \delta}(r) = - \hbar {c^{\text{3}}}{A_{,\phi }}\left( {{\phi _{\text{b}}}} \right){\iint \rho _\delta }\frac{1}{{2\pi }}{K_0}(r'/{\mathchar'26\mkern-10mu\lambda _c})d\xi d\eta, \label{ex9}
\end{equation}
where $r = {({x^2} + {y^2})^{1/2}}$, ${K_0}(r'/{\mathchar'26\mkern-10mu\lambda _c})$ is the modified Bessel function with $r' = \sqrt {{{\left( {x - \xi } \right)}^2} + {{(y - \eta )}^2}}$, and ${\rho _\delta } = ({\rho _{{\text{cyl}}}} - {\rho _{\text{b}}})\Theta ({R_{{\text{cyl}}}} - r)$ with $\Theta (x)$ being the Heaviside step function, ${\rho _{{\text{cyl}}}}$ and ${R_{{\text{cyl}}}}$ being the matter density and the radius of the cylindrical source, respectively. Eq. (\ref{ex8}) together with Eq. (\ref{ex9}) describe the fifth force generated by the cylindrical source that is embedded in the background medium in the case of $|\rho_{\delta}| \ll \rho _{\rm{b}}$. For the larger density difference of $|\rho_{\delta}|$, the linear superposition principle behaving like Eq. (\ref{ex9}) is no longer valid (the detail can be found in \cite{zhc1}).

In our VFTOF scenario, only an average acceleration of the test particle can be measured. To describe the expression of the average acceleration concisely, we introduce a function ${C(\vec r)}$ of radial direction ${\vec r}$ as follows
\begin{equation}
C(\vec r) \equiv C(x,y) = \iint {\Theta ({R_{{\text{cyl}}}} - {{({\xi ^2} + {\eta ^2})}^{1/2}})}{K_0}(r'/{\mathchar'26\mkern-10mu\lambda _c})d\xi d\eta. \label{exn10}
\end{equation}
Suppose that the test particle falls freely toward the cylindrical source along the vertical $y$ direction and the initial location of the particle is $L\,\,\rm{unit}$ away from the center axis of the source, then the average scalar fifth force along the vertical direction from $(0,L)$ to $(0,0)$ can be defined as follows
\begin{equation}
\bar a = \frac{{\hbar {c^5}}}{{2\pi }}\frac{{A_{,\phi }^2\left( {{\phi _{\text{b}}}} \right)}}{{A\left( {{\phi _{\text{b}}}} \right)}} \cdot ({\rho _{{\text{cyl}}}} - {\rho _{\text{b}}})\frac{{C(0,0)}-{C(0,L)}}{L}. \label{exn11}
\end{equation}

\begin{figure}
\centering
\includegraphics[width=200pt]{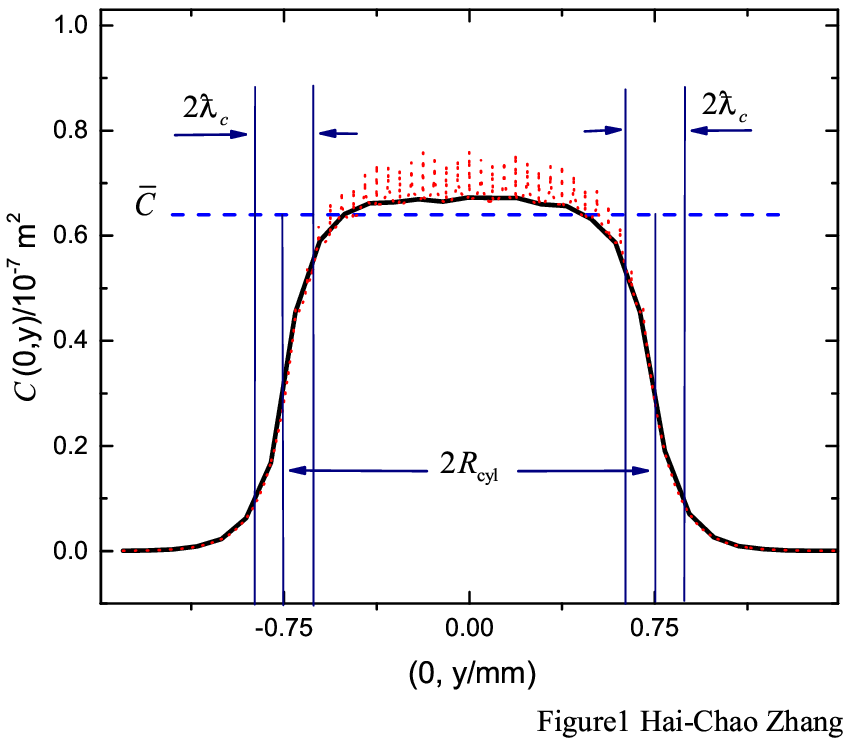}
\caption{The vertical profile of the scalar field generated by a cylindrical source. The dotted curve is obtained by the definition of Eq. (\ref{exn10}) together with the closed Universe parameters shown in Eq (\ref{ex5}), which is noisily in the inside of the source due to the numerical calculation. The solid curve is the smooth result of the dotted curve. The dash straight line corresponds to the scalar field's average value in the region of the source, which is calculated by Eq. (\ref{exn12}). The interaction range ${\mathchar'26\mkern-10mu\lambda _{\rm{c}}}$ is strongly dependent on the EBD. The larger the EBD, the shorter the Compton wavelength. Thus, in order to clearly display the character of the scalar field, we draw the curves by choosing the EBD with an imaginary value of ${\rho _{\rm{b}}} = 1 \times {10^{ - 28}}\,\rm{kg \cdot {m^{ - 3}}}$ that is even lower than the current cosmic density $\sim {10^{ - 27}}\,\rm{kg \cdot {m^{ - 3}}} $. The source radius $ R_{{\rm{cyl}}}$ is selected as $ 0.75 \, {\rm{mm}} $. }\label{figurex1}
\end{figure}

Figure \ref{figurex1} shows the function $C(0,y)$ varies along with the vertical direction of $y$ axis. In the numerical calculation, the radius of the cylindrical source is selected as an actual experimental parameter $R_{{\rm{cyl}}}= 0.75 \, {\rm{mm}}$. However, the EBD is chosen with an imaginary extremely smaller value of ${\rho _{\rm{b}}} = 1 \times {10^{ - 28}}\,\rm{kg \cdot {m^{ - 3}}}$ so as to obtain a relatively larger Compton wavelength ${\mathchar'26\mkern-10mu\lambda _{\rm{c}}}$ for drawing the illustrated graph. Besides, the calculation parameters of the curves are selected as shown as Eq. (\ref{ex5}) rather than Eq. (\ref{ex4}) because the former corresponds to a relatively large interaction range. Since the distance $L $ is often much bigger than the Compton wavelength of the scalar field, the value of $C(0,L)$ in Eq. (\ref{exn11}) is negligible. Obviously, the numerical simulation of $C(0,y)$ in the source region fluctuates a little violently although it should in deed be a smooth function of $y$. In order to avoid the uncertainty of the value of $C(0,0)$ in Eq. (\ref{exn11}), we replace $C(0,0)$ by a slightly smaller average value which is defined by
\begin{equation}
\bar C = \frac{{\int\limits_{ - {R_{{\text{cyl}}}}}^{{R_{{\text{cyl}}}}} {C\left( {0,y} \right)dy} }}{{2{R_{{\text{cyl}}}}}}.\label{exn12}
\end{equation}
In the VFTOF experiment, it is a good approximation that using $\bar C$ to replace $C(0,0)$ to calculate the acceleration of the test particle since the Compton wavelength is considerably smaller than the radius of the source. Thus, the average acceleration shown by Eq. (\ref{exn11}) can be approximately rewritten as
\begin{equation}
\bar a  \approx \frac{{\hbar {c^5}}}{{2\pi }}\frac{{A_{,\phi }^2\left( {{\phi _{\text{b}}}} \right)}}{{A\left( {{\phi _{\text{b}}}} \right)}} \cdot \frac{{({\rho _{{\text{cyl}}}} - {\rho _{\text{b}}})\bar C}}{L}.\label{exn13}
\end{equation}

From Fig. \ref{figurex1}, one sees that the fifth force is rather larger near the source surface than the other regions since it is proportional to the gradient of the scalar field. Therefore, the experimental setting of the VFTOF is designed to allow test particle can pass through the surface of source so as to sense the largest fifth force. Besides this advantage, the scenario can also enhance the experimental contrast by measuring the two different cases in the same experimental environment, that is, the density ${\rho _{{\text{cyl}}}}$ of the source is either larger or less than the EBD. When ${\rho _{{\text{cyl}}}} > {\rho _{\text{b}}}$ ( ${\rho _{{\text{cyl}}}} < {\rho _{\text{b}}}$), the source attracts (pushes away) the test particle. Therefore, by using the greatest contrast between the two cases, the existence of the fifth force may be revealed. This relative measurement is much easier to achieve a judgment for the new phenomena than an absolute measurement that needs to measure the acceleration precisely and accurately. If the fifth force exists with the extremely short Compton wavelength, it would be very difficult to measure its absolute values precisely and accurately.

To see the EBD-dependence of the fifth force intuitively, we rewrite approximately Eq. (\ref{exn13}) as follows
\begin{equation}
\bar a \propto \frac{{({\rho _{{\text{cyl}}}} - {\rho _{\text{b}}})}}{{{\rho _{\text{b}}}^2}}\frac{{\bar C}}{L}.\label{exn14}
\end{equation}
The value of ${\bar C}$ is strongly dependent on the Compton wavelength ${\mathchar'26\mkern-10mu\lambda _{\rm{c}}}$ and then on the EBD. The larger the EBD, the smaller the value of ${\bar C}$. Consequently, one can conclude that the absolute value of acceleration ${\bar a }$ decreases rapidly with the EBD growing. If $|{\rho _{{\text{cyl}}}} - {\rho _{\text{b}}}| \ll {\rho _{\text{b}}}$, the acceleration increases linearly with the microscopic matter density of the source. Notice that one cannot increase the magnitude of the fifth force further by increasing the microscopic matter density of the source when the condition $|{\rho _{{\text{cyl}}}} - {\rho _{\text{b}}}| \ll {\rho _{\text{b}}}$ is not satisfied. When the microscopic matter density of the source is large enough, the fifth force saturates at a limited value since the scalar field inside the source approaches a fixed number $M_{2}c^{2}$ \cite{zhc1}.

It is worth noting that the contribution of dark matter should in principle be counted into ${\rho _{\text{b}}}$ although it has been and will be ignored in the experimental designing for simplicity (the detail is discussed in Appendix \ref{appA}). However, the density difference of $\rho_{\delta}$ is independent of the contribution of dark matter. $\rho_{\rm{\delta}} $ is only related to atomic matter and can be generated by laser beams shown in Sec. \ref{DESIGN}.

\subsection{Summary}\label{trappedquinsummary}

When trapped quintessence is required to account for the cosmic acceleration entirely, the matter-coupled scalar field must generate the short-range fifth force due to the inhomogeneous distribution of matter including dark matter. The broken-symmetry coupling between the scalar and matter can confine the scalar at an almost fixed value and then the corresponding value of the self-interaction potential of the scalar mimics the cosmological constant. The spacial variation of the microscopic matter density results in the spacial variation of the scalar field and then the non-zero gradient of the scalar field presents the fifth force. As a typical example, a cylindrical source for generating the fifth force is analyzed specifically. The force appears only near the surface of the cylindrical source. The source will attract (repel) the test object if the matter density of the source is larger (smaller) than that of the background medium. With the background matter density growing, both the interaction range and the magnitude of the scalar fifth force decrease rapidly. The scalar field couples to non-relativistic matter but not to photons. These properties of the scalar field can be used to design the detecting experiments in laboratories on the Earth and to analyze experimental data for searching the fifth force if it really exists.

\section{OUR PROBING STRATEGY AND EXPERIMENTAL DESIGN}\label{DESIGN}

One sees that the fifth force appears only in the extremely thinner shell around the surface of the source. Accordingly, not only the source object but also the test object are designed to be constructed by the atoms of the dilute atomic gas in the VFTOF experiment, so that the every single test atom in the non-interaction atomic gas can pass through the source's surface to sense independently the largest value of the fifth force in the thin-shell.

\begin{figure}
\centering
\includegraphics[width=250pt]{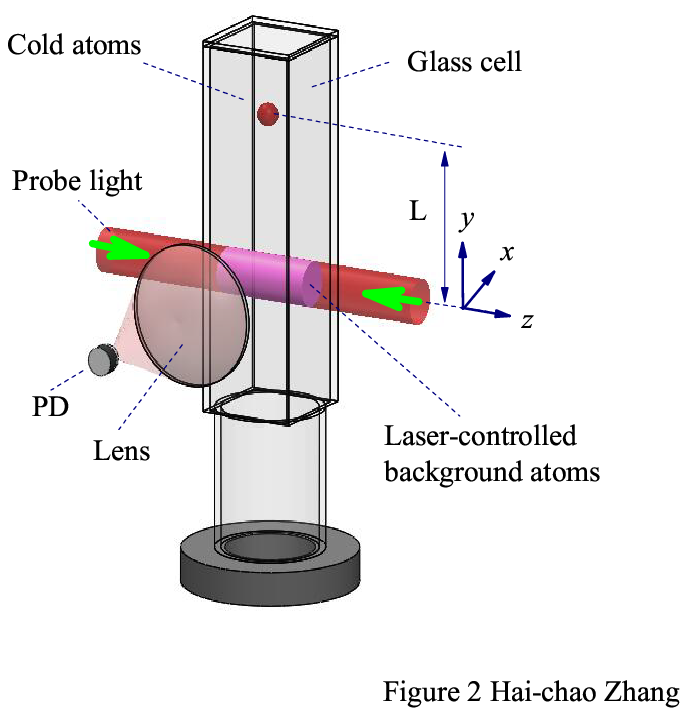}
\caption{Schematic of experimental setup for testing the scalar fifth force. The number density of background rubidium-87 atoms in the vacuum cell ($30{\text{ mm }} \times 30{\text{ mm }} \times 70 {\text{ mm}}$) was adjusted to a desired pressure by running the electric current through a dispenser (not shown). The source for generating the scalar fifth force was a part of the background atoms which was irradiated by a pair of probe laser beams. The \emph{microscopic} matter density of the source was adjusted by tuning the frequency of the probe laser light. The test atomic cloud was formed by a standard laser cooling technique, which was located $8.5{\text{ mm}}$ above the axis center of the probe light. The fall acceleration of the cold atomic cloud was measured by TOF method. If the fifth force of the source was large enough, the measured acceleration should reflect the scalar fifth force besides the gravity of the Earth. The TOF signals were recorded with a photodiode (PD) by collecting the fluorescence of the test atoms when they encounter the probe light. The lens was used to collect a lager solid angle range of the fluorescence.}\label{figurex2}
\end{figure}

Figure \ref{figurex2} shows the schematic of the experimental setup for probing the local mechanical effect of dark energy scalar field. The glass cell is an optically accessible ultrahigh vacuum chamber. The source object of the fifth force was composed of the laser-controlled background atoms. The initial matter density of the source was the same as the EBD and then changed via the energy exchange between the probe light and the atoms of the source. The detail of the source will be described in Sec. \ref{source}. The test object was composed of cold atoms that were formed by a standard laser cooling technique, which will be described in Sec. \ref{testatoms}. The fall accelerations of the test atoms were measured with the VFTOF method which will be described in Sec. \ref{resulttof}.

If the fifth force generated by the source can be sensed via the VFTOF experiment, the fall acceleration of the every single test atom will include two parts as follows: one corresponds to the gravity of the Earth; another corresponds to the fifth force. In practice, the test object in every single TOF experiment was a cold atomic cloud, which was so thin that it can be regarded as a non-interaction atomic ensemble. Thus, through a single TOF experiment with many cold atoms, one can derive an equivalent result of successive single-atom testing experiment repeated many times based on the principle of ensemble detection.

\subsection{Matter density adjustable source  }\label{source}

When the background atomic system in the ultrahigh vacuum chamber is irradiated by a pair of laser beams, there exists a net energy transfer between the light field and the atoms. Thus, the matter density adjustable source for generating the fifth force is formed in the irradiated region. By tuning the laser frequency, the matter density of the source can be adjusted to be denser or thinner than the EBD.

Conceptually, as mentioned in Sec. \ref{forceintrappedquin} the energy variation of the source is not the total energy gain or loss of the atomic ensemble in the irradiated region, but just the fractional part of the total energy gain or loss. The fractional part corresponds to the energy variation of the atoms with extension wavefunctions overlapped in the space of the irradiated region. The other much more atoms with great locality behave like classical ``billiard balls". Of course, these ``billiard-ball-like" atoms in principle can also be described by quantum mechanics. However, their wavepackets do not ``overlap" spatially, which results in a large empty space among the atoms. In other words, compared with the large empty space, the spatial volume occupied by the localized wavepackets can be safely neglected.

Consequently, when the test atoms pass through the light irradiating region, they have very little probability to encounter the localized wavepackets. Meanwhile, since the Compton wavelengths in the denser matter densities of the localized wavepackets are much smaller, the effect of the scalar field that induced by the localized wavepackets can be sensed very little by the test atoms. The test atoms can only sense the spatial variation of the scalar field that generates from the variation of the \emph{microscopic} matter density of the ``overlapping" extension wavefunctions. This is the reason that we introduce the EBD of ${\rho_{\rm{b}}}$ in Sec. \ref{forceintrappedquin} and distinguish it from the \emph{macroscopic} matter density of ${\rho_{\rm{c}}}$.

\subsubsection{\label{energyex} The estimation of the energy transfer between atom and light field }

Experimentally, the configuration of generating the matter density adjustable source is just the same as a one-dimensional (1-D) optical molasses \cite{zexn10}. There are lots of literature in calculating the energy exchange between atom and light \cite{zexn10,zhce17,zexn11}. For the sake of simplicity, we apply the two energy-level model for atom to estimate the energy transfer. The total optical dissipative force exerted on an atom in the two counterpropagating laser beams is \cite{zexn10,zhce17}
\begin{equation}
F = \hbar k({R_ + } - {R_ - }), \label{ex11}
\end{equation}
with the photon scatter rates from the two laser beams being
\begin{equation}
{R_ \pm }({v_z}) = \frac{\Gamma }{2}\frac{{I/{I_{{\rm{sat}}}}}}{{1 + 4{{\left( {\Delta  \mp k{v_z}} \right)}^2}/{\Gamma ^2} + I/{I_{{\rm{sat}}}}}}, \label{ex12}
\end{equation}
respectively. Here ${v_z}$  is the atom velocity projection onto the $z$ axis being one of the laser propagating directions, $k$ is the light wave vector of the laser beams, $\Gamma $ is the spectral width of the atom, $\Delta$ is the detuning of the laser light frequency from the atomic resonance frequency, $I_{{\rm{sat}}}$ is the saturation intensity, and $I =I_0 \exp \left( -{2{r^2}}/{\sigma _{\rm{p}}^2} \right)$ is the intensity of the optical light per beam with $I_{0}$ being the peak value of the intensity and $\sigma _{\rm{p}}$ being the $1/{e^2}$ Gaussian radius.

The energy changed per unit time is \cite{zexn10,zhce17}
\begin{equation}
W\left( {{v_z}} \right) \sim  - F{v_z}. \label{ex13}
\end{equation}
The characteristic time of the laser irradiated atomic gas approaching to equilibrium status is \cite{zexn11}
\begin{equation}
{\tau _{{\rm{ex}}}} = \frac{\hbar }{{{E_r}}},\label{ex14p}
\end{equation}
where ${E_r} = {\hbar ^2}{k^2}/(2M)$ is the recoil energy with $M$ being the atomic mass. Consequently, the mean value of the density of the transferred energy to the background atoms with overlapping extension wavefunctions is
\begin{equation}
\varepsilon_{\delta} \sim \iiint {{\tau _{{\text{ex}}}} \cdot W\left( {{v_z}} \right)f\left( {{v_x},{v_y},{v_z}} \right)}d{v_x}d{v_y}d{v_z}. \label{ex15}
\end{equation}
Here $f\left( {{v_x},{v_y},{v_z}} \right)$ is the distribution function of the background atoms with overlapping extension wavefunctions (for detail, see Appendix \ref{appB}). Thus, the variation of the microscopic mass density of the source is $\rho_{ \delta} \equiv \varepsilon_{\delta} /{c^2}$. When the microscopic mass density of non-relativistic matter is changed by the laser light, the short-range scalar fifth force of the source would appear. The laser light itself has nothing to do with the scalar fifth force since the scalar field does not couple to photons.

Now that wavefuction is involved, the quantum-statistical description of atomic motion is necessary. However, the effective interaction time $\tau _{{\rm{ex}}}$ of the optical force acting on an atom is much greater than the time scale of $\Gamma ^{-1}$. $\Gamma ^{-1}$ is the characteristic time of the process of the atom absorption followed by spontaneous emission. Thus, the atomic velocity can be regarded as a smoothly changing variable \cite{zexn11}. Although the EBD is related to the quantum property of wavefunctions, the energy transfer still can be roughly estimated by the kinetic description of classical atomic motion \cite{zexn11plus}. Notice that the atomic velocity here is indeed the group velocity of atomic wavepacket. In the language of quantum mechanics, it represents the expectation value of velocity operator in a quantum state of a single atom. The expectation value is often called the average velocity. For the same value of an average velocity, the possibilities of the width of wavepacket in the velocity-space range from zero to infinity in the nonrelativistic quantum mechanics (Of course, the width as well as the average velocity indeed cannot exceed the speed of light.). Therefore, for an atomic gas, the distribution function of the background atoms with overlapping extension wavefunctions can be obtained as follows (Appendix \ref{appB}):
\begin{equation}
f\left( {{v_x},{v_y},{v_z}} \right) = {n_{\text{b}}}{\left( {\frac{M}{{2\pi {k_B}T}}} \right)^{3/2}} \cdot \exp \left[ { - \frac{{M\left( {v_x^2 + v_y^2 + v_z^2} \right)}}{{2{k_B}T}}} \right], \label{ex16}
\end{equation}
with ${k_{\rm{B}}}$ the Boltzman constant, $n_{{\text{b}}} = {\rho _{\rm{b}}}/M$ the atomic number density corresponding to the EBD of ${\rho _{\rm{b}}}$, $T$ the temperature of the background atom gas. Then the source's microscopic mass density $\rho  = {\rho _{\rm{b}}} + \rho_{ \delta}$ can be adjusted by the detuning $\Delta$ through the detuning-dependent $\rho_{ \delta} \equiv \varepsilon_{\delta} /{c^2}$ shown as Eq. (\ref{ex15}).

The upper and lower limits of the integral on ${v_x}$ and ${v_y}$ in Eq. (\ref{ex15}) are naturally from positive to negative infinity. However, due to the counterpropagating laser beams along with the $z$ axis, the upper and lower limits of the integral on ${v_z}$ need to be analyzed further. The variation of the atomic velocity is $\hbar k/M$ during the time $\Gamma ^{-1} $ of a single ``absorption + spontaneous emission" process. Then the variation rate of the atomic velocity is $ \Gamma \hbar k/M$. The initially resonant atom would be no longer resonant with the light field after the time $\tau _{{\rm{ex}}}$ defined by Eq. (\ref{ex14p}). Thus, the upper limit of the absolute value of the variation of ${v_z}$ during the effective interaction time $\tau _{{\rm{ex}}}$ can be approximately estimated as $\tau _{{\rm{ex}}} \Gamma \hbar k/M \equiv 2\Gamma/k $. Notice that the EBD is related to the overlapping extension wavepackets. It is the wide wavepackets rather than the narrow wavepackets that can overlap easily in the space. Thus, atoms with the small velocities are more important to the EBD based on the de Broglie relations of $\lambda_{\text{de}}=h/(Mv)$. Besides, since the laser beams counter propagate along with the $z$ axis, the atoms having velocities ${v_z}$ around zero are indeed involved greatly in the process of the energy exchange between the laser light and the atoms of the source. Accordingly, the upper and lower limits of the integral on ${v_z}$ in Eq. (\ref{ex15}) are chosen as $\pm 2\Gamma/k $.

\subsubsection{\label{detuning} The detuning dependence of the laser adjusted mass density of the source}

\begin{figure}
\centering
\includegraphics[width=200pt]{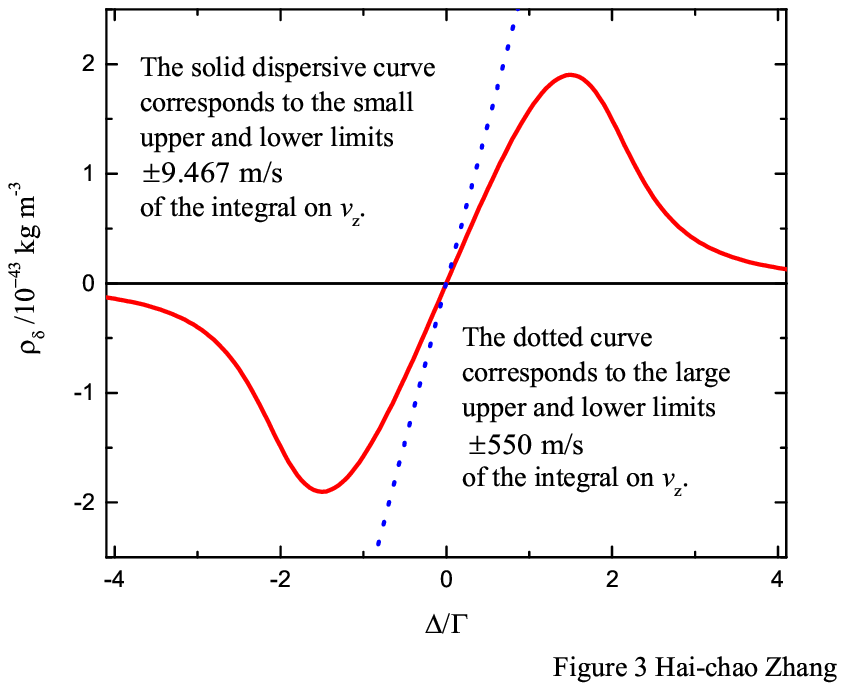}
\caption{The variation of the source's mass density as a function of the laser detuning. In order to highlight the detuning-dependence of the energy density of the source, the average intensity $\bar I = {P_0}/\left( {\pi \sigma _{\text{p}}^{\text{2}}} \right)$ rather than the spacial position dependent intensity $I$ shown as Eq. (\ref{explus9}) is used to calculate the variation of the energy density with Eq. (\ref{ex15}). The solid curve corresponds to the upper and lower limits of the integral on ${v_z}$ being $\pm  2\Gamma/k = \pm  9.467 \,\rm{m /s} $ for the $D_2$ transition of $\rm{^{87}Rb}$ atom. To recognize the importance of $2\Gamma/k $, the dotted curve is drawn for comparison, in which the upper and lower limits of the integral on ${v_z}$  are chosen as $ \pm  550 \,\rm{m /s} $ being about twice of the average velocity $270 \,\rm{m /s} $ of the atomic gas at room temperature. In both curves, the EBD is chosen as the same as ${\rho_{\text{b}}} = 1 \cdot {10^{ - 25}}{\text{ kg}} \cdot {{\text{m}}^{ - 3}}$. }\label{figurex3}
\end{figure}

Figure \ref{figurex3} shows the variation of the source's mass density versus the detuning of the light frequency from the atomic resonance frequency. The variation $\rho_{ \delta} \equiv \varepsilon_{\delta} /{c^2}$ is calculated by Eq. (\ref{ex15}). For the simplicity in illustrating the detuning-dependence of the mass density, the spacial Gaussian intensity of the light per beam in Eq. (\ref{ex12}) is replaced by an average intensity $\bar I = {P_0}/\left( {\pi \sigma _{\text{p}}^{\text{2}}} \right)$ with ${P_0}$ being the laser power per beam and $\sigma _{\rm{p}}$ being the $1/{e^2}$ Gaussian radius. The curves are calculated with the parameters of the $D_2$ line of $\rm{{}^{87}Rb}$ as follows \cite{zhce14}: $\Gamma  = 2\pi  \cdot 6.0666{\text{ MHz}}$, $M = 1.44316 \cdot {10^{ - 25}}{\text{ kg}}$, and ${I_{{\text{sat}}}} = 3.577 \,\rm{mW/cm^2}$. The parameters of the laser beams for the calculation are: the $1/{e^2}$ Gaussian radius $\sigma _{\rm{p}}=0.75 \,\rm{mm}$ , the light wave vector $ k \equiv 2\pi /{\lambda _{\rm{laser}}} $ with $ {\lambda _{\rm{laser}}} = 780.24{\text{ nm}} $ and the laser power ${P_0} = 30 \, \mu {\text{W}}$ per beam. The EBD is chosen as ${\rho_{\text{b}}} = 1 \cdot {10^{ - 25}}{\text{ kg}} \cdot {{\text{m}}^{ - 3}}$ in order to correspond to the experimental results that will be presented in Sec. \ref{RESULT}.

One sees that the mass density of the source increases when the driving laser frequency is tuned above the atomic resonance, i. e. $\Delta  > 0$, and vice versa. Thus, when $\Delta > 0$ ($\Delta < 0$), a test object would be pulled toward (pushed away from) the center of source due to the scalar fifth force generated by the source. Although the absolute value of the source density changes slightly, the fifth force would switch its direction when the sign of the detuning is shifted. Correspondingly, if the scalar field really exist, this switching property of the fifth force around zero detuning can be used to improve the contrast for probing it.

\subsubsection{\label{spaced} The radial variation of the laser adjusted mass density of the source}

We now discuss the spatial variation of the source's mass density in the radial direction ${\vec r}\equiv (x,y)$ when the two laser beams counter propagate along with the $z$ axis. Assume that the intensity of the per laser beam has a Gaussian profile as
\begin{equation}
I\left( x ,y \right) = {I_0}\exp \left( { - \frac{{{2 x ^2} + {2 y ^2}}}{{{\sigma _{\rm{p}}}^2}}} \right), \label{explus9}
\end{equation}
where the peak value of the intensity ${I_0} = {2{P_0}/\left( {\pi \sigma _{\rm{p}}^2} \right)}= 2\bar I$, then the spatial profile of the laser adjusted mass density can be obtained by Eq. (\ref{ex15}) together with the relation of $\rho_{ \delta} \equiv \varepsilon_{\delta} /{c^2}$. The repeated mention of the mass-energy relation is to emphasize that the fifth force can be produced only when the light energy is transformed into non-relativistic mass.
\begin{figure}
\centering
\includegraphics[width=200pt]{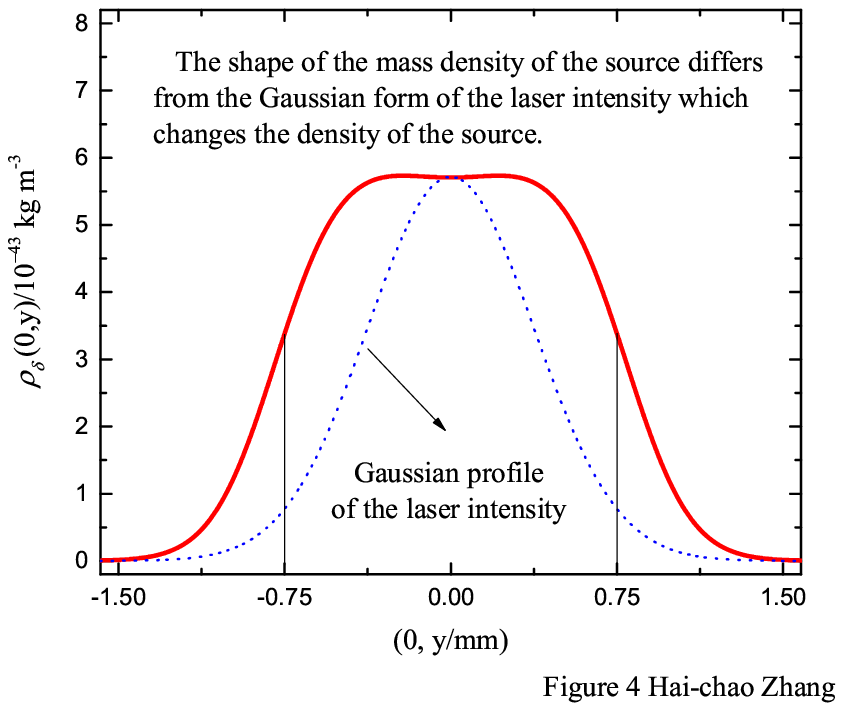}
\caption{The spatial profile of the laser adjusted matter density of the source. The solid curve shows that the matter density varies along with the $y$ axis, which is numerical calculated by Eq. (\ref{ex15}) with the spatial dependent intensity $I$ shown by Eq. (\ref{explus9}) in a fixed value of the detuning. The parameters in the calculation are: the per beam power ${P_0} = 250 \, \mu {\text{W}}$, the $1/{e^2}$ Gaussian radius $\sigma _{\rm{p}}=0.75 \,\rm{mm}$ of the laser beam, the value of detuning selected as $\Delta=1\cdot \Gamma $, and the EBD selected as ${\rho_{\text{b}}} = 1 \cdot {10^{ - 25}}{\text{ kg}} \cdot {{\text{m}}^{ - 3}}$. The dotted curve shows the Gaussian profile of the laser intensity for comparing to the profile of the mass density of the source. The profile of the mass density of the source does not behave like the Gaussian profile of the laser intensity and indeed depends on the laser power. }\label{figurex4}
\end{figure}

Figure \ref{figurex4} shows that the profile of the mass density of the source (the solid curve) is different from the Gaussian form of the laser intensity (the dotted curve). In the case of a low laser intensity (${I \ll I_{{\text{sat}}}}$), form Eq. (\ref{ex12}), the excited state population of the two-level system for atom is almost proportional to the laser intensity. Correspondingly, the shape of the source matter density profile is almost the same as that of the laser intensity profile. However, in the case of the actual experiment, the intensity value in the center of laser intensity profile is often larger than that of the saturation density (${I_{0} > I_{{\text{sat}}}}$), whereas in the wings of the profile, it may be not. Thus, the absorption of light in the central region of the profile is often saturated, whereas in a fixed position of the wing regions, the saturation parameter of the absorption depends on the actual value of laser power. Consequently, the profile of the source matter density is more complicated than that of the laser intensity. By choosing appropriate parameters of the laser, the spatial variation of the source's mass density along with the radial direction can behave approximately like a cylinder rather than a Gaussian shape. Therefore, in order to avoid the complicated calculation of the fifth force, the laser adjusted source will be regarded approximately as a cylinder in the following text. Notice that, even in this case, the profile of the cylinder-like source varies with the laser power. The size of the source density profile will be broadened when the laser power increases.

\subsubsection{\label{forceofsource} The fifth force of the laser adjusted source }

Let us now estimate the order of the average fifth force shown by Eq. (\ref{exn13}) with the same parameters as Fig. \ref{figurex3}. In order to compare with the VFTOF experiment in Sec. \ref{RESULT}, the other parameters that is needed to calculate the force are chosen as follows: $L=8.5\, \rm{mm}$, $R_{\rm{cyl}}=\sigma _{\rm{p}}=0.75\, \rm{mm}$. Since the fifth force is strongly dependent on the spatial curvature of the Universe, we calculate the fifth force in the two cases shown in Eqs. (\ref{ex4}) and (\ref{ex5}), respectively.

\begin{figure}
\centering
\includegraphics[width=200pt]{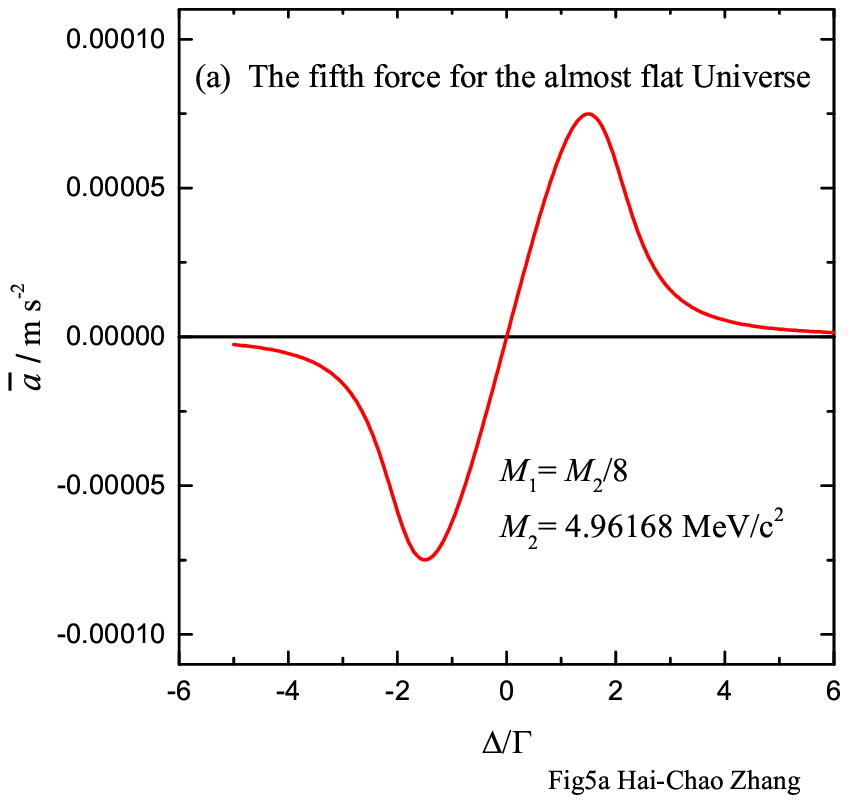}
\includegraphics[width=190pt]{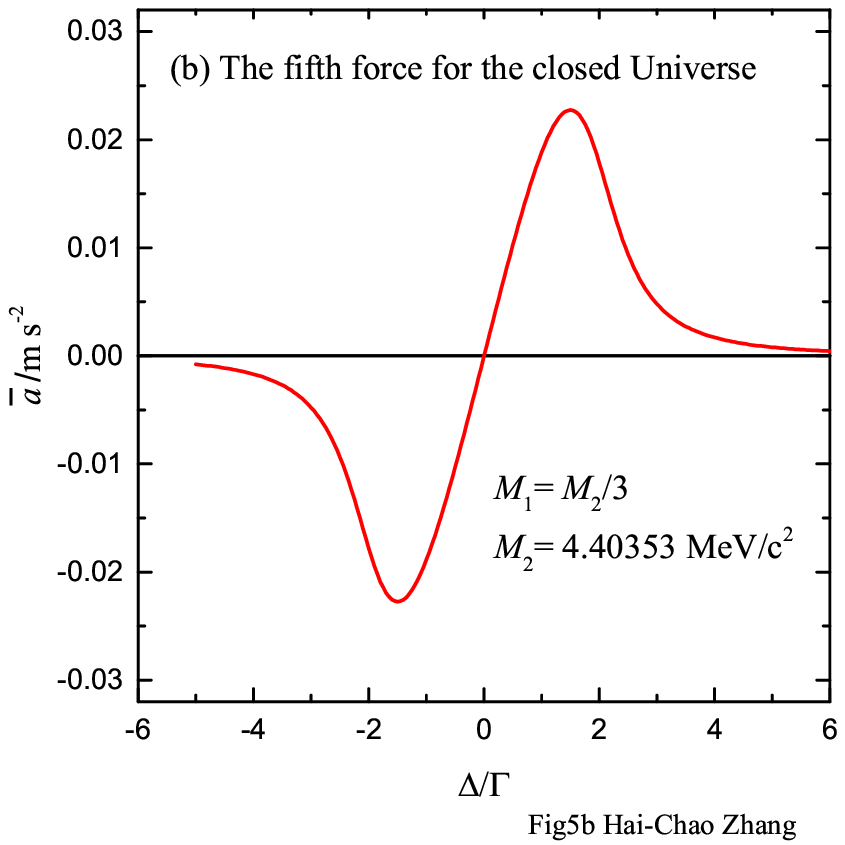}
\caption{The scalar fifth force of the laser adjusted source versus the light frequency detuning from the atomic resonance frequency. The magnitude of the fifth force depends on the spatial curvature of the Universe. (a) Using the parameters of $M_1$ and $M_2$ in the almost flat Universe shown in Eq. (\ref{ex4}); (b) Using the parameters of $M_1$ and $M_2$ in the closed Universe shown in Eq. (\ref{ex5}). The mass density of the source is regarded as a cylindrical form of ${\rho  } = ({\rho _{{\text{cyl}}}} - {\rho _{\text{b}}})\Theta ({R_{{\text{cyl}}}} - r)+{\rho _{\text{b}}}$, where the radius of the cylinder is selected as the $1/{e^2}$ Gaussian radius of the laser beam, i.e., $R_{\rm{cyl}}=\sigma _{\rm{p}}=0.75\, \rm{mm}$. The difference between the source's mass density and the EBD is calculated using Eq. (\ref{ex15}) with the average light intensity $\bar I = {P_0}/\left( {\pi \sigma _{\text{p}}^{\text{2}}} \right)$. The initial distance $L$ between the test atomic cloud and the source is selected to be $8.5\, \rm{mm}$. In both curves (a) and (b), the EBD and the laser power per beam are the same as that in Fig. \ref{figurex3}, i.e., ${\rho_{\text{b}}} = 1 \cdot {10^{ - 25}}{\text{ kg}} \cdot {{\text{m}}^{ - 3}}$, ${P_0} = 30 \, \mu {\text{W}}$, respectively. }\label{figurex5}
\end{figure}

Figure \ref{figurex5} shows the detuning dependence of the average fifth force. One sees that the contrast for probing the fifth force can be enhanced by comparing the experimental results in the either cases of the negative and positive detuning (see also Appendix \ref{appA}). One also sees that the fifth force in the closed Universe would be bigger than that in the almost flat Universe. Thus, the effect of the spatial curvature may be detected determinately in the future by improving the measurement technique of the scalar fifth force.

It should be emphasized that the fifth force would emerge in the outside of the source due to the finite interaction range of the scalar field. This property assures that the test object can not only sense the fifth force outside the source but also avoid the disturbance of the laser beams inside the source. If the Compton wavelength is extremely short so that the test object cannot distinguish the region of the fifth force from that of the laser beams, the existence of the fifth force is not able to be demonstrated firmly in the VFTOF experimental configuration even if it really exists.

\subsection{The requirement for test atoms and their preparation method }\label{testatoms}

In order to sense the fifth force generated by the laser adjusted source, the test atoms should be prepared to be able to reach the surface of the source as close as possible due to the fifth force being large in the neighbourhood of the surface. In fact, the test atoms in the VFTOF configuration can even pass through all the region of the source. When the test atoms approach the surface of the source, they would be speeded up (slowed down) if the microscopic matter density of the source is larger (smaller) than the EBD. The test atoms sense firstly the fifth force in the outside region of the source and then encounter the laser beams. The atoms will fluoresce when they arrive at the region of the laser beams. Therefore, it is very important that the test object can feel sensitively the fifth force before it encounters the laser beams. Or else, it would be difficult to judge whether the scalar fifth force exists or not.

One of the best ways to test the scalar fifth force in the VFTOF configuration may be the method of imaging a single test atom in each TOF experiment and repeating the experiment many times as shown in literature \cite{zhcn15}. However, based on the principle of ensemble detection, we can also use many non-interaction atoms in a single TOF experiment to probe the fifth force. In this case, the initial distribution function of the atoms in phase space (which composes of configuration space and momentum space) is often assumed as a product of the two Gaussian profiles which correspond respectively to configuration space and momentum space \cite{zhce21}.

However, in a real physical situation, the atomic ensemble sample is limited by not only the interaction between atoms but also Heisenberg's uncertainty relation between the momentum and position of particles. To fulfil the condition of non-interaction between atoms, the separation between atoms should be much larger than the size of atoms. The size of atom is often characterized by the s-wave scattering length. For $\rm{^{87}Rb}$, the s-wave scattering length is about $100a_{\rm{B}}$ \cite{zhce16}, where $a_{\rm{B}}$ is the Bohr radius. In the so called condition of diluteness, the composite particles can be regarded as pointlike particles. To fulfil the classical description of the atomic gas that can be treated as a system of ``billiard balls", the width of the velocity distribution of the gas is wide enough so as not to achieve the status of degenerated quantum gas such as Bose-Einstein condensation (BEC).

Unlike the source that is based on the extension property of the atomic wavepacketes, the test atomic sample is required to be based on the localization property of the wavepackets of the test atoms. Therefore, in contrast to the extremely small value of the EBD, the non-interaction atoms in the test atomic sample can achieve a huge number. Using the standard laser cooling technique known as three-dimensional (3-D) optical molasses \cite{zexn10,zhce17,zhce18,zhce19,zhce20}, the required test atoms can be formed and suspended inside of an ultrahigh-vacuum chamber. When the test atomic cloud is released by switching off the laser beams of optical molasses, the atomic cloud not only falls vertically toward the source but also expands ballistically due to the initial distribution of atomic velocities. Thus, when the non-interaction atoms approach the surface of the source and then pass through the surface, the information of the scalar fifth force generated by the source would be sensed by the test atoms.  The detailed description of the experimental process will be shown in Sec. \ref{RESULT}.

\subsection{VFTOF method of searching the scalar fifth force }\label{resulttof}

There are a few precision experiments \cite{z102,z47,z80} including the atomic interferometry in probing the dark energy scalar fifth force. If the fifth force theorized by the trapped quintessence model \cite{zhc1}, the interaction range of the matter-coupled scalar is extremely smaller than the distances between the sources and the sensors in these experiments. Therefore, the atomic interferometry in the current configurations cannot sense the fifth force. However, in the TOF configuration, the test atoms can pass through the surface of the source and then can sense the largest fifth force in the thin-shell. One cannot deduce the existence of the fifth force by the TOF experiment only at a fixed value of the detuning between the laser and atomic frequencies since there is large systematic error in measuring the distance between the probe laser and the initial position of the cold atomic cloud. Measuring the acceleration of the test atomic cloud via TOF method is sensitively to the distance. To eliminate the systematic error, one can use VFTOF method, i.e., by varying the frequency of the laser light step-by-step to perform a series of TOF experiments in the detuning frequency domain. More importantly, the existence of the scalar fifth force may be deduced by the contrast of the measured fall accelerations between the positive and the negative detuning.

\subsubsection{\label{TOFmethod} TOF method for measuring the fall acceleration of the test atoms }

Generally speaking, TOF method refers to a technique of deriving some properties of media by measuring the time spent by test objects (including particles and waves in a medium) over a certain distance. In our scheme, the cold atoms are used as the test objects. The test atoms will fluoresce when they encounter the counter-propagating laser beams that are used to change the mass density of the source firstly. The TOF signals are obtained by collecting the fluorescence of the test atoms. Therefore, the counter-propagating laser beams are called probe light as marked in Fig. \ref{figurex2}.

From Eq. (\ref{exn13}), the average fifth force is inversely proportional to the distance $L$ between the initial position of the test atomic cloud and the location of the probe light. Thus, in order to obtain a larger value of the average fifth force, the distance of $L$ should be as short as possible. The TOF signal at time $t$ in the case of short distance is \cite{zhce21}

\begin{equation}
S\left( t \right) \propto \frac{{{P_0}}}{{\sigma _0^2 + \sigma _v^2{t^2} + \sigma _{\rm{I}}^{\rm{2}}}}\exp \left\{ { - {{\left( {\frac{{{ a_{\rm{fall}}}\left( {t_0^2 - {t^2}} \right)}}{{2\sqrt 2 \sqrt {\sigma _0^2 + \sigma _v^2{t^2} + \sigma _{\rm{I}}^{\rm{2}}} }}} \right)}^2}} \right\}, \label{ex22}
\end{equation}
where ${\sigma _0}$ is the initial $1/\sqrt e $ Gaussian radius of the test atomic cloud; ${\sigma _v}$ is the $1/\sqrt e $ Gaussian radius of the velocity distribution of the atom cloud, which associated its temperature by the formula of ${T_{{\rm{test}}}} = m\sigma _v^2/{k_{\rm{B}}}$; $ a_{\text{fall}}$ is the fall acceleration of the atomic cloud; ${t_0} = \sqrt {2L/{ a_{\rm{fall}}}} $ is the arrival time of the center of the test atom cloud without initial vertical velocity; $\sigma _{\rm{I}}$ is the $1/\sqrt e $ Gaussian radius of the laser beams and $\sigma _{\rm{I}}=\sigma _{\rm{p}}/2$ due to $\sigma _{\rm{p}}$ being the $1/{e^2}$ Gaussian radius. The above radii are described by using $1/\sqrt e $ instead of $1/{e^2}$ in order to respect the original literature \cite{zhce21}.

By fitting experimental TOF signal with Eq. (\ref{ex22}), one can derive the fall acceleration of the test atomic cloud. The fall acceleration should be a sum of
\begin{equation}
{a_{{\text{fall}}}} = g + \bar a, \label{ex23}
\end{equation}
where $g$ is the gravitational acceleration of the Earth and $\bar a$ is the scalar fifth force shown by Eq. (\ref{exn13}).

Apparently, when TOF experiments are performed only at a fixed value of the laser detuning, even if the measured acceleration of the cold atomic cloud is very different from the gravity acceleration $g$, it cannot be completely accepted to attribute this difference to the fifth force. The reason is the possibility of the existence of the systematic errors. Thus, introducing a new method to eliminate the systematic errors is necessary.

\subsubsection{\label{commethod} The comparison approach to eliminate the systematic error }

Since the fifth force generated by laser adjusted source is an anti-symmetric function of the detuning as shown in Fig. \ref{figurex5}, we can measure the values of the fifth force by VFTOF method to obtain the experimental curve in the frequency domain of the detuning. If the experimental curve behaves like the profile shown in Fig. \ref{figurex5}, it may be accounted for the scalar fifth force theorized by the trapped quintessence model \cite{zhc1}. The judgments of the existence of the dark energy scalar by the comparison approach between both cases of the positive and the negative detuning is independent of the systematic errors of the measurements of the related distance and time. For example, if a slight larger value of the distance $L$ is used to fit the TOF experimental signals, according to Eq. (\ref{ex22}) and the definition ${t_0} = \sqrt {2L/{ a_{\rm{fall}}}} $ , all the fitted values of the fall accelerations in the detuning frequency domain would be larger than their true values. But the shape of the dispersion curves with respect to the detuning should remain almost unchanged if the dispersion shape is observable.

Conversely, the distance $L$ can be taken as a fitting parameter to fit the dispersion curves of the measured accelerations with respect to the detuning. Thus, the systematic error caused by the bias of the distance is eliminated. The fitted value of the distance $L$ should be feed back to the fitting process of the TOF signals in order to derive the new fall accelerations in the the detuning frequency domain. The measuring accuracy of the scalar fifth force is then improved.

The fluorescence of the background atoms induced by the probe light may also exerts radiation pressure to the cold atomic cloud \cite{zhce22}. However, unlike that the fifth force is an anti-symmetric function of the detuning as shown by Fig. \ref{figurex5}, the radiation pressure generated by the fluorescence of the background atoms (RPFB) is a symmetric function of the detuning. Therefore, for avoiding the complexity, we have deliberately abandoned the contribution of the RPFB to the fall acceleration ${a_{{\text{fall}}}}$ as shown by Eq. (\ref{ex23}). This omission of the symmetric function of the detuning will not affect our judgment of the existence of the fifth force since the judgment will be based on the anti-symmetric function of the detuning.

\subsection{Summary}\label{designsummary}

The source for generating the fifth force to be measured can be formed by the space-overlapping extension wavepackets of the background atoms in an ultrahigh vacuum chamber. The overwhelming majority of the background atoms behave like the localized wavefunctions and the corresponding microscopic matter density is strongly peaked at atom locations but approaches zero in the large void regions between atoms. Thus, the microscopic matter density of the source is extremely smaller than the macroscopic matter density that corresponds to the background vacuum degree. The microscopic matter density of the source can be slightly adjusted by laser light. The slight variation of the matter density of the source can result in an observable fifth force based on the trapped quintessence model of dark energy.

The characteristic of very short interaction-range of the scalar requires us to design such an experiment that the sensor can approach and cross the surface of the source. The cold atomic cloud formed by the optical molasses technique can serve as a sensor of the fifth force generated by the source. In fact, the analysis of the mechanical effect of scalar field on the test atoms is based on the ballistic motion of the overwhelming majority of the atoms in the cold atomic cloud. The typical TOF configuration for cold atoms can satisfy the above experimental requirements for probing the scalar fifth force. The theoretical curve of the fifth force versus the detuning between the laser and atomic frequencies is an anti-symmetric function of the detuning. Based on the VFTOF method, the dispersion curve may be observed in laboratories on the Earth if the trapped quintessence model is correct.

\section{EXPERIMENTAL PROCEDURE AND RESULTS }\label{RESULT}

In this Section, we will show how to demonstrate experimentally the detuning-dependence of the scalar fifth force by the VFTOF method. The experimental setup shown in Fig. \ref{figurex2} can be easily replicated in laboratories on the Earth. The detailed description of the experimental procedure can be found in \cite{zhce22}.

\subsection{The TOF experimental procedure }\label{exproce}

The TOF experiments with $\rm{^{87}Rb}$ atoms were performed in the ultrahigh vacuum chamber as shown in Fig. \ref{figurex2}. The background rubidium atoms in the vacuum chamber were obtained by running a current through a rubidium dispenser. The background gas pressure of the vacuum chamber was adjusted to be $\sim3.5\times10^{-8} \,\rm{Pa}$, which corresponds to the \emph{macroscopic} number density $n_{\text{c}} \sim 10^{13} \,\rm{m^{-3}}$ of $\rm{^{87}Rb}$ atoms, or equivalently, the \emph{macroscopic} mass density of ${\rho_{\text{c}}} \sim {10^{ - 12}}{\text{ kg}} \cdot {{\text{m}}^{ - 3}}$. It is the microscopic matter density rather than the \emph{macroscopic} matter density that determines the motion of the scalar field. The EBD ${\rho_{\text{b}}}$ of the background atoms is just the smooth part of the microscopic matter density of the background atoms. The lumpy part of the microscopic matter density is peaked at the locations of the localized wavepackets but is zero in the large spaces between the localized wavepackets. In general, the \emph{macroscopic} matter density can be regarded as an average density of the microscopic matter density over a certain \emph{macroscopic} scale. The \emph{macroscopic} matter density cannot be used to describe any feature of sources that generate the scalar fifth forces since it cannot reflect the fact of the large almost empty spaces between the localized atoms.

Any single atom has very little probability not only to encounter a localized atom in the ultrahigh vacuum chamber, but also to sense the fifth force generated by the localized atom since the interaction-range of the scalar is extremely short. Consequently, the source that generates the fifth force to be measured in the VFTOF scheme does not refer to the all atoms in the region covered by the probe light. It corresponds only to the probe-laser-adjusted EBD.

The variation of the mass density of the source was adjusted by tuning the frequency of the probe light. The probe light was formed by two counter-propagating laser beams which were made by using a beam splitter \cite{zhce19}. The two counter-propagating beams were used to change the matter density of the source of the fifth force, as well as to act the probe light in the VFTOF method. The laser beams with $1/e^2$ Gaussian radius $\sigma_{\rm{P}}= 0.75\,\rm{ mm}$ were tuned to a series of fixed detuning values around the $\rm{5S_{1/2}}$ to $\rm{5P_{3/2}} $ cycling transition $F_{g}=2\rightarrow F_{e}=3$. Since both the power of the probe light per beam and the detuning between the probe light and the atomic frequencies are frequently addressed, we have marked them as ${P_0}$ and $\Delta$, respectively. The two counter-propagating laser beams also contain a certain power of re-pumping light (which was locked to some detuning values from the $F_{g}=1\rightarrow F_{e}=2$ resonant transition) to prevent the atom from being pumped to a lower ground state hyperfine structure level. Since both the power of the re-pumping light per beam and the frequency of the re-pumping light in the probe light were needed to be changed in the measurement process, we mark them by $P_{\rm{repump}}$ and $\Delta_{\rm{repump}}$, respectively. If only one laser beam rather than the two counter-propagating laser beams is used to excite the source and to act the probe light, the atom will absorb photons in one direction for many times to obtain momentum. This will cause the atom leave out the probe region and the analysis of the energy transfer in the last Sec. \ref{source} is no longer valid.

We used cold atomic cloud to sense the scalar fifth force generated by the source. The cold atoms of $\rm{^{87}Rb}$ were pre-prepared firstly in a conventional magneto-optical trap (MOT) \cite{zhce18}. The trapping beams ($1/e^2$-intensity contour diameter $ 7.5\,\rm{mm}$ for each beam; power $6.8\,\rm{mW}$ per beam) were detuned by $13\,\rm{ MHz}$ below the $\rm{5S_{1/2}}$ to $\rm{5P_{3/2}} $ cycling transition $F_{g}=2\rightarrow F_{e}=3$. The repumping laser was locked to $F_{g}=1\rightarrow F_{e}=2$ resonant transition to prevent optical pumping of the atoms to the lower ground-state hyperfine level. The MOT was loaded about $6$ seconds. After the loading process, the atomic cloud was further cooled by optical molasses \cite{zexn10,zhce17}. In general, the prepared test atoms after the molasses process have atomic number $N_{\rm{mol}}$ on the order of $\sim  {10^7}$ with temperature $T_{\rm{mol}} $ of the order on tens $ {\rm{ \mu K}}$. The initial $1/\sqrt e $ Gaussian radius of the atom cloud after the molasses process was determined to be as ${\sigma _0} = 0.8{\text{ mm}}$ by fluorescence imaging with a charge-couple-device camera (CCD). The typical mass density $\rho_{\rm{mol}}$ of the cold atomic cloud was estimated to be on the order of $\sim 10^{-10} {\text{ kg/}}{{\text{m}}^3}$, corresponding to the number density $n_{\rm{mol}}$ on the order of $\sim 10^{15}\,\rm{m^{-3}}$. Thus, the distance between the cold atoms is on the order of several $\rm{\mu m}$, which is greatly larger than the size of the $\rm{^{87}Rb}$ being on the order of several $\rm{nm}$. The distance between the test atoms is also greatly larger than the thermal de Broglie wavelength $h/(2\pi k_{\rm{B}} T_{\rm{mol}}M)^{1/2} \sim10\,\rm{ nm}$ where $h$ is the Planck constant. Consequently, the vast majority of the test atoms behave like non-interaction distinguishable classical particles and then satisfy the requirement of ensemble detection principle for probing the fifth force generated by the source. The molasses process was performed about ten millisecond before the cooling beams were shut off and the cold atomic cloud began to free fall and expand.

The probe laser light was switched on at the beginning when the MOT was shut on so as to assure that the energy exchange between the light field and the background atoms achieved an equilibrium status completely. The mass density of the source is defined in the equilibrium state. The initial time of the TOF signals began at the time when the molasses was shut off, which marks that the fall acceleration measurement was started. We monitored the fluorescence of the test atoms excited by the probe beams with a PD to obtain TOF signal. To collect the fluorescence from a larger solid angle range, a lens was used as shown in Fig. \ref{figurex2}. It should be emphasized again that the probe beams were endowed with twofold functions: one is for adjusting the mass density of the source that produces the fifth force; the other is for exciting the test atoms to produce fluorescence when the atoms encounter the probe beams. By fitting the TOF signals, the fall acceleration of the atomic cloud was obtained.

\subsection{The detuning dependence of the measured fall accelerations }\label{datafitting}

In order to use the comparison method to judge whether the measured accelerations behave like the theoretical predictions as shown in Fig. \ref{figurex5}, the TOF measurement was performed point-by-point in the frequency domain of the laser detuning. The detuning $\Delta  = {\omega _{{\text{laser}}}} - {\omega _{{\text{atom}}}}$ is the difference between the probe light frequency ${\omega _{{\text{laser}}}}$ and the atomic resonance frequency ${\omega _{{\text{atom}}}}$. During the experimental process, the data acquisition was adopted in alternating mode of a series of positive and negative detuning to eliminate the systemic error in the detuning frequency domain. The detailed measurement process by the VFTOF method is described as follows.

\begin{figure}
\centering
\includegraphics[width=185pt]{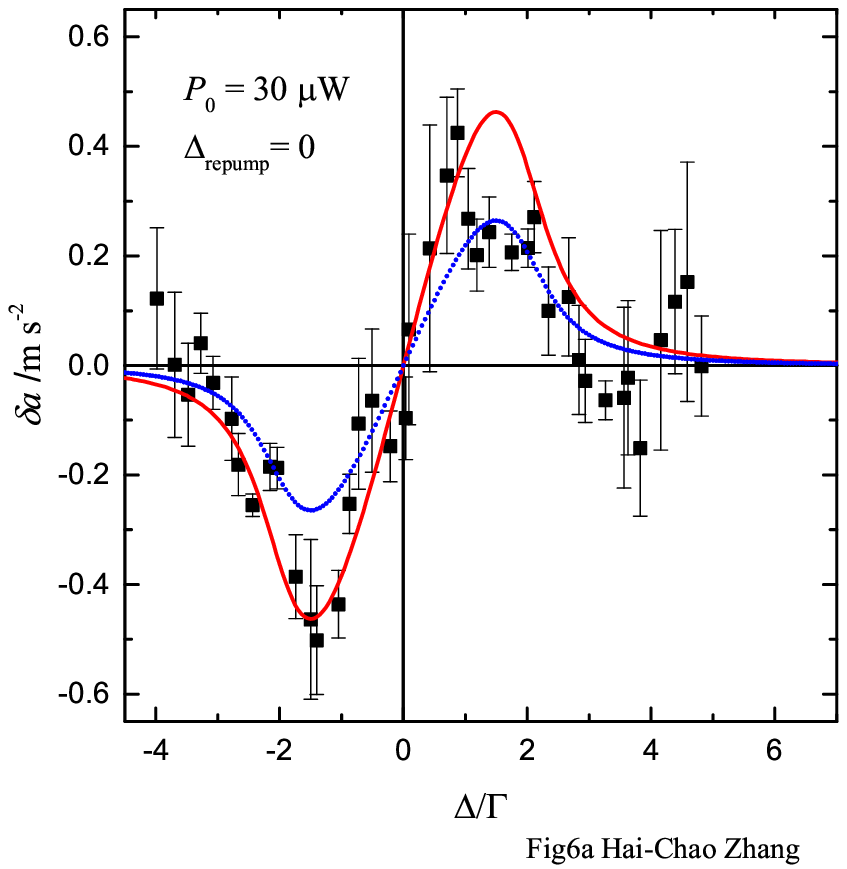}
\includegraphics[width=200pt]{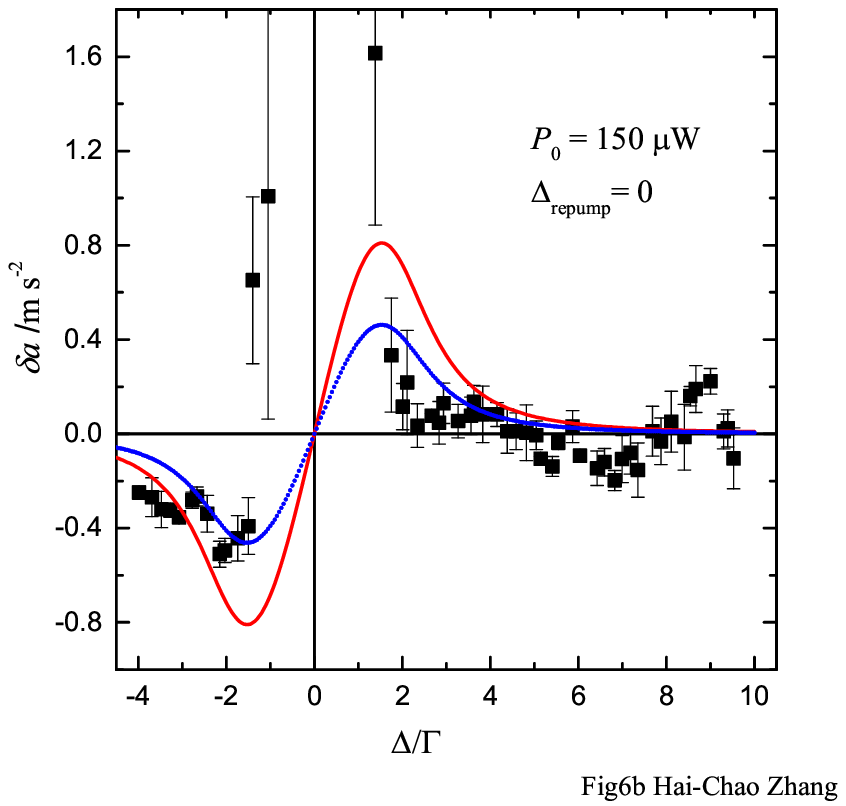}
\includegraphics[width=200pt]{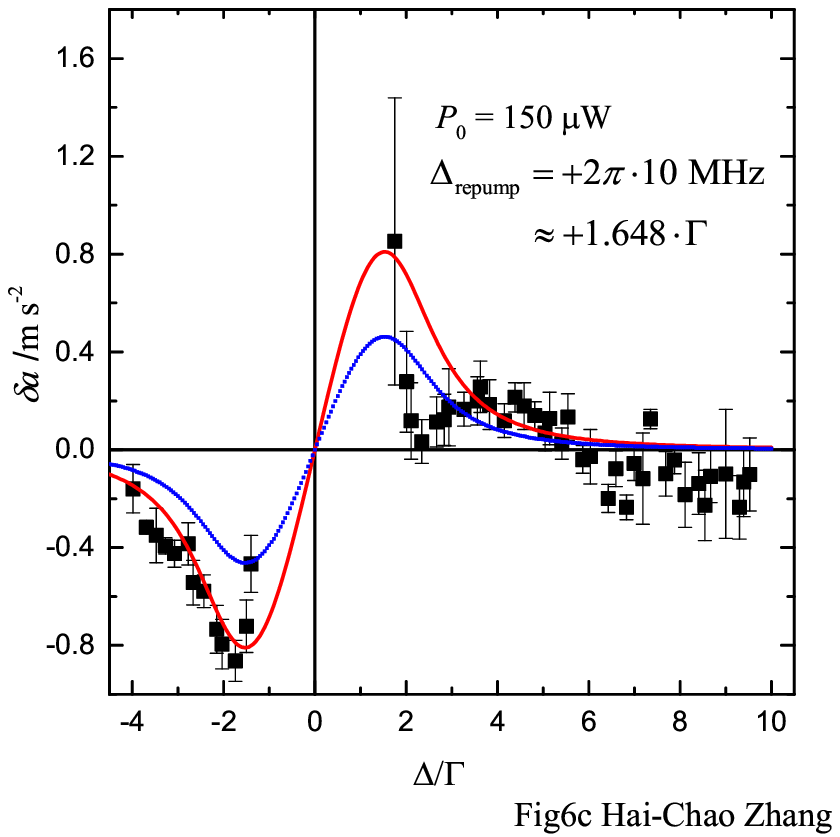}
\includegraphics[width=200pt]{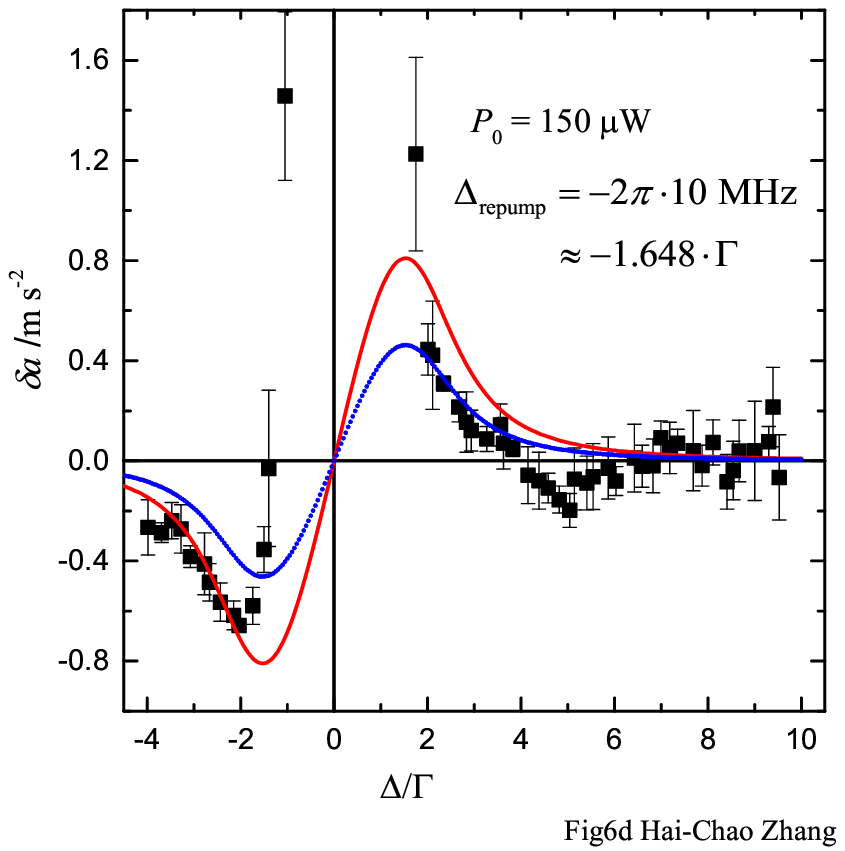}
\caption{The acceleration variations versus the detuning of the probe beams. The experimental data are marked by the squares. Every square corresponds to an average value over the four values of the acceleration that are fitted from the four times repeated TOF signals. The error bar for each square is the standard deviation based on the four measurements. (a) The probe power per beam was ${P_0} = 30\,{\rm{ \mu W}}$ and the re-pumping light was locked to the $F_{g}=1\rightarrow F_{e}=2$ resonant transition, i.e., $\Delta_{\rm{repump}}=0$. (b) The probe power per beam was ${P_0} = 150\,{\rm{ \mu W}}$ and the re-pumping light detuning was $\Delta_{\rm{repump}}=0$. (c) The probe power per beam was ${P_0} = 150\,{\rm{ \mu W}}$ and the re-pumping light detuning was $\Delta_{\rm{repump}}= +2\pi \cdot 10\,\rm{MHz}$. (d) The probe power per beam was ${P_0} = 150\,{\rm{ \mu W}}$ and the re-pumping light detuning was $\Delta_{\rm{repump}}= -2\pi \cdot 10\,\rm{MHz}$. The re-pumping power per beam in the probe beams was fixed as $P_{\rm{repump}}=300\, {\rm{ \mu W}}$ for all the cases of (a), (b), (c) and (d). The calculated curves are derived by using Eq. (\ref{exn13}) together with Eq. (\ref{ex15}) based on the assumption that the experimental data $\delta a$ can be explain as the scalar fifth force shown in Eq. (\ref{exn13}). The solid curves correspond to the EBD ${\rho_{\text{b}}} = 1.3 \cdot {10^{ - 26}}{\text{ kg}} \cdot {{\text{m}}^{ - 3}}$. The dotted curves correspond to the EBD ${\rho_{\text{b}}} = 1.9 \cdot {10^{ - 26}}{\text{ kg}} \cdot {{\text{m}}^{ - 3}}$. In all of the calculated solid and dotted curves, the cosmic parameters are selected as the parameters for the closed Universe, i.e., $ M_{2} = 4.40353{\text{ meV/}}{c^2}$ and $ {M_1} = {M_2}/3$ shown in Eq. (\ref{ex5}). If we use the parameters for the almost flat Universe as shown in Eq. (\ref{ex4}) as $ {M_2} = 4.96168 \; {\rm{meV/}}{c^2}$ and $ {M_1} = {M_2}/8$, the EBD ${\rho_{\text{b}}}$ is fitted on the order of $ \sim 5 \cdot {10^{ - 27}}{\text{ kg}} \cdot {{\text{m}}^{ - 3}}$ being slightly larger than or near to the current cosmic density. The two probe beams here were linearly and parallel polarized. The similar experimental dispersion curves can also be demonstrated in the case of circular polarized probe beams.}\label{figurex6}
\end{figure}

Varying the frequency of the probe light to change the detuning while fixing the power of the probe light, we measured the fall accelerations of the test atom clouds under different detuning. The probe power per beam was selected as two fixed values as ${P_0} = 30\,{\rm{ \mu W}}$ and ${P_0} = 150\,{\rm{ \mu W}}$, respectively. The $1/{e^2}$ Gaussian radius of the probe beams was ${\sigma _{\text{p}}} = 0.75{\text{ mm}}$. The distance between the center of the MOT and the center axis of the probe beams was $L=8.5\, \rm{ mm}$. In order to ensure that the experimental conditions are almost the same for positive and negative detuning, the TOF signals were recorded alternately between positive detuning and negative detuning, beginning from both ends of the maximum positive detuning and the maximum negative detuning. Under the same experimental conditions, the TOF signal acquisition experiment was repeated four times at every detuning point. To highlight the detuning-dependence part of the fall acceleration, we introduce an acceleration variation $\delta a$ which is defined by the difference between the measured fall acceleration and the average acceleration that was obtained by averaging the data points for the large detuning of $\Delta  \gtrsim 3 \Gamma $ for each data group. Using the definition of the average over the detuning range $\Delta  \gtrsim 3 \Gamma $ implies an assumption that for large detuning the fall acceleration will not vary with the detuning. The choice of detuning range, such as $\Delta\gtrsim 3\Gamma$, does not need to be too strict, because the acceleration variation $\delta a $ is introduced only for the convenience of description. The acceleration variations $\delta a$ at the different detuning $\Delta$ are shown as squares in Fig. \ref{figurex6}.

From Fig. \ref{figurex6}, one sees that the fitted values of the fall accelerations in the red-detuning region are smaller than that in the blue-detuning region. Although the two probe beams here were linearly and parallel polarized, the similar dispersion curves of the accelerations can also be obtained in the case of circular polarized probe beams. One cannot regard that the detuning-dependence may results from some feature of the polarization of the probe beams. The dispersion property of the fitted acceleration versus the detuning \emph{does not} depend on the polarization of the probe light.

Since the probe laser beams were located under the cold atomic clouds in the experimental setup, the fluorescence of the laser excited background atoms could only lower rather than heighten the fall accelerations \cite{zhce22}. The RPFB symmetric function with respect to the axis of $\Delta =0$ cannot be used to explain the detuning dependence of the fall accelerations since the experimental fitting data form the approximate anti-symmetric functions of the detuning.

We also cannot attribute the detuning-dependent accelerations to the optical dipole potential well that might be formed by the probe laser beams. In fact, the intensity of the probe light is too weak to form an observable gradient force to affect the cold atoms. Even if the gradient force is observable, it gives a reversal effect compared to the experimental results, i.e., for the red-detuning the optical dipole force drives the cold atoms to positions where the light intensity has a maximum, whereas for the blue-detuning the optical dipole force pushes the atoms away from the intensity maximum.

\subsection{\label{explanationdetuningdependen} Using the trapped quintessence model to explain the measured dispersion curves }

The detuning-dependent acceleration of the cold atomic cloud cannot be explained with our limited knowledge in traditional physics. Because the experimental curves of $\delta a$ versus $\Delta$ can be numerically simulated by Eq. (\ref{exn13}) together with Eq. (\ref{ex15}), the origin of the detuning-dependent acceleration may be accounted for the scalar field coupled with matter via the symmetry-breaking interaction \cite{zhc1}. The scalar with self-interaction potential density is originally invoked to mimic the cosmological constant to drive the accelerated expanding of the Universe. The symmetry-breaking coupling between matter and the scalar must result in the scalar fifth force and then leads to the fall accelerations of the test atoms varying with the laser detuning. The detailed analysis of the measured dispersion curves of the accelerations with the trapped quintessence model is given as follows.

The solid curves in Figs. \ref{figurex6}(a)-\ref{figurex6}(d) are calculated by Eq. (\ref{exn13}) with the EBD ${\rho_{\text{b}}} = 1.3 \cdot {10^{ - 26}}{\text{ kg}} \cdot {{\text{m}}^{ - 3}}$, which are coincided basically with the experimental data in the red-detuning region. The dotted curves in Figs. \ref{figurex6}(a)-\ref{figurex6}(d) are calculated by Eq. (\ref{exn13}) with the EBD ${\rho_{\text{b}}} = 1.9 \cdot {10^{ - 26}}{\text{ kg}} \cdot {{\text{m}}^{ - 3}}$, which are in general close to the experimental data in the blue-detuning. The cosmic parameters in the calculations are chosen as $ M_{2} = 4.40353{\text{ meV/}}{c^2}$ and $ {M_1} = {M_2}/3$ shown in Eq. (\ref{ex5}) which correspond to the closed Universe.

Obviously, although the theoretical curves are similar to the experimentally measured acceleration dispersion curves with respect to the detuning, the calculated results cannot be adjusted to be completely consistent with the experimental data in the whole detuning region. This tension can be explained as a systematic error in measuring the distance $L$ between the initial position of the atom and the probe light. The indirect measurement of the fall acceleration of the test atom is extremely sensitive to the deviation of the measured value of the distance $L$.

The current experimental data shown in Fig. \ref{figurex6} are obtained by fitting the TOF signals with the value of the distance $L=8.5 \,\rm{mm}$. Theoretically, $L$ is the distance between the center of the cold atomic cloud and the axisymmetric center of the probe light. However, it is very difficult to define and locate the center of the cold atom cloud accurately. The reasons are given as follows. In every experimental process, the shape of the cold atomic cloud was neither symmetrical nor regular. Since the TOF technique is a destructive measurement for the cold atomic cloud, the atomic cloud needs to be prepared again in the next time during the repeated experiments. The shapes and sizes of these cold atomic clouds cannot be kept exactly the same in repeated experiments. Therefore, the value of the distance was derived actually by measuring the distance between the two axis centers of the probe beams and the horizontal cooling beams which form one of three pairs of cooling light. Since the diameter of cooling beams was large as $ 7.5 \,\rm{mm}$, the axis center of the cooling beams was located by using a variable aperture in the place where the beam splitters have not been used to form the six cooling beams. The center of the variable aperture was well adjusted to approach the axis center of the cooling beams. If the two centers of the aperture and the axis of the cooling light were aligned well with each other, the MOT can still be observable by CCD even if the aperture became very small. However, since the optical molasses is related to six cooling beams, the effective center of the cold atomic cloud in the actual experimental situation may have a little deviation from the optical measurement of $L=8.5 \,\rm{mm}$.

Consequently, we can take the distance $L$ as a fitting parameter to obtain new experimental data, so that the new experimental data are agreement with a calculated result in the whole detuning region (not shown by graph to avoid confusion with the original experimental data shown in Fig. \ref{figurex6}). By using Eq. (\ref{exn13}) together with Eq. (\ref{ex15}) and based on the cylindrical source, the value of the distance $L$ is fitted to be about $8.9 \,\rm{mm}$. Besides, the EBD is fitted to be as a median value of ${\rho_{\text{b}}} = 1.6 \cdot {10^{ - 26}}{\text{ kg}} \cdot {{\text{m}}^{ - 3}}$ which happens to be the average of the two values $ 1.3 \cdot {10^{ - 26}}{\text{ kg}} \cdot {{\text{m}}^{ - 3}}$ and $ 1.9 \cdot {10^{ - 26}}{\text{ kg}} \cdot {{\text{m}}^{ - 3}}$ mentioned above. Thus, if the theoretical model of trapped quintessence is correct, its prediction of the anti-symmetric fifth force with respect to the laser detuning can even be used to infer the actual value of the distance between the center of MOT and probe beams based on the assumption that experimental data of $\delta a$ equal to the scalar fifth force of $\bar a$ shown in Eq. (\ref{exn13}).

Interestingly, the fitted value of the distance $L$ does not depend on the choice of the cosmic parameters shown in Eq. (\ref{ex5}) or Eq. (\ref{ex4}). This means that the actual value of distance is indeed independent of the spatial curvature of the Universe as one would expect. It is the estimating deviation of the distance that leads to the systematic error of the above measurements for the accelerations of the cold atomic clouds. However, the fitted value of the EBD is very sensitive to the spatial curvature of the Universe. If we choose the cosmic parameters as $ {M_2} = 4.96168 \; {\rm{meV/}}{c^2}$ and $ {M_1} = {M_2}/8$ shown in Eq. (\ref{ex4}) which correspond to the almost flat space of the current Universe, then the EBD ${\rho_{\text{b}}}$ is fitted to be $ \sim 5 \cdot {10^{ - 27}}{\text{ kg}} \cdot {{\text{m}}^{ - 3}}$ which is slightly larger than or near to the current matter density of the Universe. If the fitted value of the EBD is very smaller than the cosmic density, especially, the dark matter density, the corresponding cosmic parameters may be questionable since fuzzy dark matter with ultralight mass is supposed to fill the cosmic space everywhere in the trapped quintessence model. The spatial curvature of our Universe cannot be determined by using the experimental data in this paper. However, with the improvement of the measurement precision, the effects of the spatial curvature of the Universe may be observed in laboratories on the Earth in the near future if the trapped quintessence model is correct.

In contrast to the dispersive curve shown in Fig. \ref{figurex6}(a), there are sharp elevations of the acceleration distributions near zero detuning of $\Delta =0$ in Figs. \ref{figurex6}(b)- \ref{figurex6}(d). In short, the appearance of the heightened effect of the fall acceleration near $\Delta =0$ for the large power of ${P_0} = 150\,{\rm{ \mu W}}$ is due to the addition of a very small number of cold atoms to the source that generates the scalar fifth force. If there are no cold atoms in the source, there will be no such protuberances. The detailed analysis is given as follows.

Since the probe power of ${P_0} = 150\,{\rm{ \mu W}}$ is $5$ times of ${P_0} = 30\,{\rm{ \mu W}}$, for the same absolute value of laser intensity contour, the effective action scale of the laser beam with the former power is larger than that of the laser beam with the latter power due to the Gaussian shape of the intensity of $I =I_0 \exp \left( -{2{r^2}}/{\sigma _{\rm{p}}^2} \right)$. Although the both probe powers correspond to the same $1/e^2$-intensity contour diameter, the laser intensity at the former probe beam's $1/e^2$-intensity contour is $5$ times of that at the latter probe beam's $1/e^2$-intensity contour. Thus, the probe light with larger power was more likely to encounter a few of cold atoms cooled by the cooling beams. These cold atoms could be heated by the probe light in the resonance situation but could not be cooled further by the probe light due to the extremely low temperature of the cold atoms. This means that the energy of the source for producing the scalar fifth force must increase since there must be cold atoms with overlapping wavepackets to contribute to the source. Therefore, the source behaved like to attract the test atomic cloud and then the fall accelerations of the test atoms were enhanced. This phenomenon further shows that it is non-relativistic matter rather than radiation that interacts with the scalar field. Unfortunately, it is difficult to estimate the value of the fifth force when the source contains some cold atoms with overlapping extension wavepackets, because we do not know the exact energy transferred to the hybrid source from the laser light.

The heightened effect of the fall acceleration near $\Delta =0$ for the large power of ${P_0} = 150\,{\rm{ \mu W}}$ cannot be explained by the RPFB since the RPFB can only decrease the fall acceleration in the current experimental configuration \cite{zhce22}. To suppress the heightened effect near $\Delta =0$, probe light should be located far enough from the region of the cooling laser beams so as to avoid cold atomic ingredients mixing into the source that generates the fifth force. However, if the distance between probe the light and the MOT is too large, it would be difficult to observe the dispersion behaviour of the fall acceleration of cold atomic cloud versus the detuning of the probe light, because the average acceleration shown in Eq. (\ref{exn13}) is inversely proportional to the distance.

\subsection{Summary}\label{datasummary}

The fall acceleration of the cold atomic cloud was measured by using the TOF method. The measured acceleration was strongly dependent on the detuning between the probe laser and the atomic resonance frequencies. Because the gravitational acceleration of the Earth has nothing to do with the laser light frequency, the measured dispersion curves of the fall acceleration with respect to the detuning are considered to be caused by the detuning-dependent scalar fifth force. The source for generating the fifth force gained (lost) positive energy when the probe laser frequency was tuned above (below) the atomic resonance, so the mass density of the source increased (decreased) compared to that of the background atomic gas. Consequently, the source attracted (repelled) the test atoms in the case of the blue-detuning (red-detuning) and then the dispersion behavior of the fall acceleration appeared.

The distance between the source and the test cold atomic cloud was selected to be large enough so as to avoid the source containing a few components of cold atoms. Otherwise, the cold-atom-doped source could be heated by the probe beams in the resonance case and then pulled the test atoms downward due to the matter density enhancement of the source. Even if the distance selected only satisfied this condition above for the lower power of the probe light, e.g., ${P_0} = 30\,{\rm{ \mu W}}$, the effective radial action range of the probe light would increase when the power of the laser light increased, e.g., ${P_0} = 150\,{\rm{ \mu W}}$. The probe light with larger power had a larger probability of encountering a few of cold atoms in the periphery of the test cold atomic cloud cooled by the cooling laser beams. Thus, the larger fall accelerations of the test atoms near zero detuning were observed in the case of large probe power of ${P_0} = 150\,{\rm{ \mu W}}$ since the source was mixed with some cold atoms.

Based on the dispersion curves measured by the VFTOF, the values of the EBD are fitted in both cases of the closed Universe and the almost flat Universe. The fitted values of the EBD in both cases are larger than the current matter density of the Universe, especially, the current dark matter density, satisfying the constraint of the trapped quintessence model which states that fuzzy dark matter needs to fill all the cosmic space. The fitted results of the EBD mean that the cosmic space curvature cannot be adjudged determinedly by using the current experimental data under the trapped quintessence model.

\section{Discussions and Conclusions}\label{Conclusions}

The trapped quintessence model of dark energy predicts a matter-density-dependent short range fifth force. The larger the matter density, the shorter the interaction range of the scalar fifth force. The extremely short range results in that the fifth force appears only in the neighborhood of the interface between the background and the source that produces the fifth force. At the interface the scalar fifth force reaches its peak value. If the matter density of the source approaches but is not equal to that of the background, the peak value is roughly proportional to the matter density difference between the source and the background but \emph{inversely} proportional to the square of the matter density of the background. If the density difference is comparable to or larger than the background density, the peak value will approach a fixed value with the density difference increasing due to the nonlinear character of the equation of motion for the scalar field.

It should be emphasized that the matter density here is the microscopic matter density which is defined by the wavepackets of particles. For the dilute atomic gas in an ultrahigh vacuum chamber, the homogeneous background density is formed by the overlapping extension wavefunctions of the atoms, which is very different from the average density in the chamber. The locally peaking part of the microscopic matter density reflects the sparsely dotted and floated individual billiard-ball-like atoms with the narrowly localized and spatially separated wavepackets in the chamber. The space volume occupied by the billiard-ball-like atoms is negligible. Except for the background formed by the overlapping extension wavefunctions of the atoms, the space between the billiard-ball-like atoms is empty. We indeed do not know exactly the ratio of the number of atoms forming the background to the total number of atoms. But we can infer that the ratio is extremely small based on the estimation of the thermal de Broglie wavelength. The ratio is experimentally demonstrated to be on the order of $10^{-14}$ by using ${\rho_{\text{b}}} / {\rho_{\text{c}}}$ under the trapped quintessence model. This ratio is cosmic-parameter-dependent in the trapped quintessence model. Thus, if the EBD of ${\rho_{\text{b}}}$ is deduced even smaller than the current dark matter density with some group cosmic parameters, then these cosmic parameters are questionable since fuzzy dark matter is needed to fill the cosmic space everywhere in the trapped quintessence model.

We constructed a matter-density-variable source by irradiating the background atoms in the ultrahigh vacuum chamber with the laser beams. In the case of ${\rho_{\text{b}}} \sim 10^{-26}\,\rm{kg/m^3}$, the change of the mass density of the source can be achieved on the order of $10^{-44}\,\rm{kg/m^3}$ due to the laser irradiation, which corresponds to the fifth force on the order of $0.1\,\rm{m/s^2}$. Such large values of the fifth force can be identified by the VFTOF experiment setup if the interaction range of the scalar is short just like several of the $\rm{\mu m's}$ in the case of ${\rho_{\text{b}}} \sim 10^{-26}\,\rm{kg/m^3}$. The main advantages in our measurement methodology are: (i) The test atoms can pass through the thin-shell of the source to sense the largest value of the force; (ii) by scanning the TOF signals point-by-point in the detuning frequency domain, the comparison method can be used to adjudge the detuning-dependent force.

The test atomic cloud for probing the fifth force were formed by the optical molasses technique. By recording the time spent by the test atomic cloud over the distance between its initial position and the probe position, the fall acceleration of the test atomic cloud can be derived by the TOF method. If there is no other force exerting on the test atoms besides the gravity of the Earth, the fall acceleration of the test atoms should not vary with the detuning of the probe light from the atomic resonance transition. However, the detuning-dependent fall acceleration of the test atoms was derived the VFTOF. We have failed to explain this dispersion phenomenon with our limited knowledge in traditional physics. We attribute the detuning-dependent part of the fall acceleration to the scalar fifth force generated by the source.

In conclusion, we demonstrated experimentally by the VFTOF method that the fall acceleration of the test atoms is dependent on the frequency of the probe light and explained this measured dispersion phenomenon with a scalar fifth force theorized by the trapped quintessence model. When the nonrelativistic matter density of the source increased due to the energy gained from the laser light, the test atoms were pulled to the center of the source, and vice versa. The fifth force in the trapped quintessence model is considerably larger than Newtonian gravity in short distances. However, the interaction range of the scalar is short enough to satisfy all experimental constraints on deviations from GR. If the trapped quintessence model is correct, the observed detuning-dependent acceleration in the VFTOF scheme suggests a closed Universe, although the exact value of the positive spacial curvature of the Universe cannot be determined in the current measurement sensitivity and accuracy. As long as the scale factor of the Universe is large enough, a closed Universe can behave like a flat one (the detail is discussed in Appendix\ref{appC}). Under the constraint of the cosmological constant, the trapped quintessence model predicts that, in the same laboratory experimental situation, the smaller the cosmic radius, the larger the scalar fifth force.

\begin{acknowledgments}
We acknowledge discussions with Jie-Nian Zhang and Jian-Da Shao. This work was supported by the National Natural Science Foundation of China through Grant No. 12074396.
\end{acknowledgments}

\appendix

\section{The role of dark matter in the experimental design and results}\label{appA}
\setcounter{section}{1}

The interaction range of the fifth force in Earth-based laboratories depends on the total microscopic density of matter including atomic matter and dark matter. However, in the literature, the value of the dark matter density is estimated to be from $0$ \cite{randr7,randr13,randr15} to $\sim 10^{-22}\,\rm{kg/m^3}$ \cite{randr17,randr9,ranr18}. To empirically measure the fifth force, we do not consider dark matter in the experiment design process and the analysis of the experiment results, although the trapped quintessence model depends on the hypothesis that the cosmic space is filled with dark matter everywhere. For relatively light dark matter particles, we can assume that the density of dark matter near the Earth approaches the current cosmic density (the reason will be discussed in the next paragraph). Therefore, if the experimental fitting value of matter density is much smaller than that of the current cosmic density, we must give a reasonable explanation for the lack of dark matter. Fortunately, our experimental fitting values are larger than the current cosmic density, which means that the light component of dark matter is safely ignored.

Now we discuss the density of dark matter on the surface of the Earth where the experiment is performed. In the trapped quintessence model, the cosmic space is filled with fuzzy dark matter. If we take the microscopic density of dark matter of the voids between galaxies on the order of $\sim 10^{-27}\,\rm{kg/m^3}$ (the current cosmic density), the dark matter density near an astronomical object, such as the Earth, can be roughly estimated by the well-known barometer formula \cite{randr25}:
\begin{equation}
\varrho \left( r \right) = {\varrho_\infty }\exp \left( { - \frac{{mU\left( r \right)}}{{{k_{\text{B}}}T_{d}}}} \right),  \label{appa1}
\end{equation}
where ${\varrho_\infty } $ refers to the dark matter density at the reference position of infinity, $m$ the mass of one particle of dark matter, $T_{d}$ the temperature of dark matter and $U(r)$ the gravitational potential of the Earth. Assume the Earth of mass $M_{e}$ to be a uniform solid sphere of radius $r_0$, the gravitational potential is well-known as follows:
\begin{equation}
U\left( r \right) =
\begin{cases}
- \frac{{G{M_e}}}{r} &  r > {r_0}\\
\frac{{G{M_e}{r^2}}}{{2r_0^3}} - \frac{{3G{M_e}}}{{2{r_0}}} & r \leq {r_0}, \label{appa2}
\end{cases}
\end{equation}
where $G$ is the gravitational constant. Due to the lack of knowledge of dark matter's temperature, the temperature value is now assumed to be taken as the current temperature of the CMB. The current temperature of the CMB is $T_{0}=2.73\, \rm{K}$ \cite{randr2}. Using Eqs. (\ref{appa1}) and (\ref{appa2}), one can see that, for $m < 1 \,\rm{meV}$, the dark matter density at the surface of the Earth is almost the same as that at the infinity reference position, i.e., $\varrho (r_{0}) \simeq {\varrho_\infty } $. Notice that the self-gravity between the particles of dark matter is ignored. Since the reduced mean thermal wavelength of $\hbar /(2\pi m k_{B} T_{d})^{1/2}$ of the particles is larger than the distance of $ (\varrho_\infty /m)^{-1/3} $ between the particles in the case of $m < 1 \,\rm{meV}$, the extension of the wavefunctions of the particles can fill the cosmic space everywhere. We know that, the smaller the particle mass, the larger extension of the wavefunction. Thus, the fuzzy dark matter of ultralight mass can completely permeate the entire cosmic space without cracks. In fact, when the mean thermal wavelength of the particles is larger than the distance between the particles, the classical barometer formula is no longer valid. We must use the Fermi-Dirac statistics for fermions or Bose-Einstein statistics for bosons to describe the quantum behavior of the system \cite{randr26}.

In our experimental situation, the atomic EBD is fitted to be $\sim 10^{-26}\,\rm{kg/m^3}$. Since we assume that the fuzzy dark matter density on the order of the current cosmic density $\sim 10^{-27}\,\rm{kg/m^3}$, the influence of the fuzzy dark matter to the fifth force can be ignored. However, it should be emphasized that, although the experimental fitting value of the atomic EBD is larger than the current cosmic density, the experiment result cannot exclude the existence of one or more heavy-mass components of dark matter. For example, due to the short interaction range of the fifth force, the experiment here cannot sense a very heavy dark matter particle which is located far away from the experiment platform. For dark matter particles with mass equivalent to atomic mass, we may also classify the particles into two types: ``billiard ball" type and wavepacket extension type to discuss the fifth force. When the EBD of dark matter is used to describe homogeneous density background of the Universe, the cosmological constant can be roughly estimated from the dark matter EBD. Of course, ``the cosmological constant" is indeed not a constant in any quintessence-like models. The variable ``cosmological constant" in space-time depends on the distribution of all the particles of dark matter and baryonic matter. The gravitational potential at any space-point is the sum of the gravitational potential of all the particles of dark matter and baryonic matter. Consequently, the following may happen in the trapped quintessence model: the number of dark matter particles required to match the cosmological constant seems to be less than that required to match the observation in galaxies, since the cosmological constant can be roughly estimated from the homogeneous part of matter distribution of the Universe.

\section{The generalized Maxwell-Boltzmann distribution function}\label{appB}
\setcounter{section}{2}

For a quantum state $\left| \phi  \right\rangle$ of a single particle, the average kinetic energy (the expectation value of kinetic energy operator) of the particle can be divided into two parts as follows:
\begin{equation}
\left\langle \phi  \right|\frac{{{{\hat p}^2}}}{{2M}}\left| \phi  \right\rangle  \triangleq \overline {\left( {\frac{{{{\hat p}^2}}}{{2M}}} \right)}  = \overline {{{\frac{{\left( {\overline {\hat p}  + (\hat p - \overline {\hat p} )} \right)}}{{2M}}}^2}}  = \frac{{{{\left( {\overline {\hat p} } \right)}^2}}}{{2M}} + \frac{{\overline {{{\left( {\Delta p} \right)}^2}} }}{{2M}}, \label{appb1}
\end{equation}
where $\hat p$ and $M$ are the momentum operator and the mass of the particle. One part of ${(\overline {\hat p} )^2}/(2M)$ corresponds to the average momentum $\bar p$ (the expectation value of momentum operator $\hat p$ in the quantum state $\left| \phi  \right\rangle $); another part of $\overline {{{\left( {\Delta p} \right)}^2}} /(2M)$ corresponds to the root-mean-square fluctuation or the uncertainty about the momentum of the particle, i.e., $\Delta p$. It is well-known that in statistical mechanics the famous Boltzmann factor $\exp(-\beta\varepsilon)$ is related to the \emph{total energy} $\varepsilon$ of a single particle \cite{randr26}. Thus, for an atomic gas, by assuming that the average velocity (the expectation value of velocity operator $\hat p/M$) distribution is independent of the velocity fluctuation distribution, the Maxwell-Boltzmann distribution function can be generalized as follows:
\begin{equation}
f\left( {{{\bar v}_x},{{\bar v}_y},{{\bar v}_z};\Delta {p_x},\Delta {p_y},\Delta {p_z}} \right) = C\exp \left( { - \tfrac{{M\left( {\bar v_x^2 + \bar v_y^2 + \bar v_z^2} \right)}}{{2{k_B}T}} - \tfrac{{{{\left( {\Delta {p_x}} \right)}^2} + {{\left( {\Delta {p_y}} \right)}^2} + {{\left( {\Delta {p_z}} \right)}^2}}}{{2M{k_B}T}}} \right), \label{appb2}
\end{equation}
where $C$ is a normalizing constant.

Now we discuss the EBD. If we use $\Delta q $ to denote the extension of the wavepacket in position-space, the formation condition of the homogeneous density distribution of the atomic system can be expressed as
\begin{equation}
\Delta q \geq {n^{ - {1/3}}},\label{appb3}
\end{equation}
where $n$ is the atomic number density. According to Heisenberg's uncertainty relation \cite{randr27}
\begin{equation}
\Delta p \cdot \Delta q = \zeta \hbar \label{appb3plus}
\end{equation}
with ${1/2} \leq \zeta  < \infty$, the formation condition of the EBD can be rewritten as follows:
\begin{equation}
\Delta p \leq {n^{{1/3}} }\zeta \hbar. \label{appb4}
\end{equation}

Consequently, by integrating the generalized Maxwell-Boltzmann distribution function (\ref{appb2}) over the $\Delta p$-space from 0 to ${n^{{1/3}} }\zeta \hbar $, the average velocity distribution of the background atoms with extension wavefunctions can be obtained as follows:
\begin{equation}
f\left( {{{\bar v}_x},{{\bar v}_y},{{\bar v}_z}} \right) = {n_{\text{b}}}{\left( {\frac{M}{{2\pi {k_B}T}}} \right)^{3/2}} \cdot \exp \left[ { - \frac{{M\left( {\bar v_x^2 + \bar v_y^2 + \bar v_z^2} \right)}}{{2{k_B}T}}} \right] \label{appb5}
\end{equation}
in which
\begin{equation}
 {n_{\text{b}}} = {\left( {\frac{1}{{2\pi M{k_B}T}}} \right)^{3/2}}\int_0^{{n^{{1/3}}}\zeta \hbar } {\exp \left( { - \tfrac{{{{\left( {\Delta p} \right)}^2}}}{{2M{k_B}T}}} \right)} 4\pi {\left( {\Delta p} \right)^2}d\left( {\Delta p} \right) \label{appb6}
\end{equation}
corresponds to the EBD, i.e., $\rho_\text{b}\equiv M{n_{\text{b}}}$. To see intuitively the dependence of $n$, $T$ and $M$, one may rewrite the definition of $n_{\text{b}}$ as follows:
\begin{equation}
{n_{\text{b}}} = \frac{4}{{\sqrt \pi  }}\int_0^{\zeta {\pi ^{1/2}}{n^{1/3}}{\mathchar'26\mkern-10mu\lambda _{{\text{th}}}}} {{e^{ - {x^2}}}{x^2}dx} , \label{appb7}
\end{equation}
where ${\mathchar'26\mkern-10mu\lambda _{{\text{th}}}} \equiv \hbar {\left( {2\pi M{k_B}T} \right)^{ - 1/2}}$ is the reduced thermal de Broglie wavelength. The average symbols in Eq. (\ref{appb5}) have been omitted in Eq. (\ref{ex16}) for brevity. It is worth noting that the average velocity here is related to a quantum wavefuntion of a single atom not to any distribution function of an atomic gas.

The value of $\zeta$ depends on the concrete physics system. When the smallest value $1/2$ of $\zeta$ is applied, the lower limit on EBD of the atomic system is estimated to be on the order of $10^{-48}\,\rm{kg/m^3}$ in our experimental condition, which is extremely smaller than the current cosmic density $\sim 10^{-27}\,\rm{kg/m^3}$. If the atomic EBD were as small as $10^{-48}\,\rm{kg/m^3}$, the contribution of dark matter should be counted into the total EBD in the calculation of fifth force. Because dark matter is not considered to be interacting with photons, the variation of the microscopic mass density of the laser-excited-source is calculated by using the distribution function of the background atoms with overlapping extension wavefunctions. In our experimental situation, the atomic EBD is fitted to be $\sim 10^{-26}\,\rm{kg/m^3}$, corresponding to $\zeta \sim 10^7$ which satisfies the requirement of Heisenberg's uncertainty relation (\ref{appb3plus}). Since the density of dark matter is estimated to be on the order of the current cosmic density (see also Appendix \ref{appA}), the fitting value of EBD is then greater than the density of dark matter. Therefore, we cannot infer more information about dark matter from the current experiment.

\section{The flatness problem in a closed Universe}\label{appC}
\setcounter{section}{3}

The flatness and horizon problems in the standard big-bang cosmology have been solved by the inflationary scenario originally proposed by Guth and Sato, and improved by Linde and Steindard et. al. (Please see \cite{randr2} for review). However, inflation does not change the global geometric nature of the Universe. In other words, if the Universe was open, closed or flat at the beginning of the era of inflation, it would still be open, closed or flat after inflation. There is no physical mechanism to convert the Universe from one spatial geometric attribute to another. Nearly all modern astronomical observations show that the Universe is almost spatially flat, with literature \cite{z114} as an exception. In discussing the flatness problem, the Friedmann equation of the Universe is often rewritten as follows \cite{randr2}
\begin{equation}
\Omega  - 1 = \frac{{K{c^2}}}{{{a^2}{H^2}}}, \label{appc1}
\end{equation}
where $\Omega$ is the density parameter denoting the ratio of the cosmic energy density to the critical density, $a$ the scale factor of the Universe, $H\equiv \dot{a}/a$ the Hubble expansion rate, $c$ the speed of light and $K =0, 1, $ or $ -1$ the normalized spatial curvature. Form Eq. (\ref{appc1}), one can see that the flatness issue of $\Omega \rightarrow 1$ can be fulfilled by the three options as follows:

(1) The normalized spatial curvature $K \equiv 0$, which corresponds to an absolutely flat space.

(2) The Hubble parameter $H$ approaches infinity, which is the original motivation of the inflation proposal to solve the flatness problem. Of course, the essential nature of inflation is that the term of ${a^2}{H^2}$ rapidly increases during the era and then leads to $\Omega \rightarrow 1$ rapidly. This does not mean that $K$ is equal to zero.

(3) The scale factor $a$ approaches infinity, which corresponds to the current issue in this paper. Unlike the inflationary period, $\Omega $ does not experience a violently variation. Due to a considerably large scale factor in the present era, $\Omega $ can remain in the status with its value being very close to 1.

Thus, a closed cosmic model with $K=1$ does not contradict the astronomical observations of spatial flatness as long as the scale factor $a$ of the Universe is large enough. The conclusion that $ K = 0 $ cannot be drawn from the observations of flatness.


\begin{thebibliography}{100}
\bibitem{zexn11plus} When the extension of the atomic wavepacket is large compared with the laser wavelength, some semiclassical approximations in the treatment of the atomic motion is no longer valid. The fully quantum treatment is necessary \cite{zexn11}.
\end{thebibliography}
\end{document}